\documentclass[journal]{IEEEtran}
\usepackage{amsmath,amsfonts}
\usepackage{algorithmic}
\usepackage{algorithm}
\usepackage{array}
\usepackage{textcomp}
\usepackage{subcaption}
\usepackage{stfloats}
\usepackage{url}
\usepackage{verbatim}
\usepackage{cite}
\def\BibTeX{{\rm B\kern-.05em{\sc i\kern-.025em b}\kern-.08em
T\kern-.1667em\lower.7ex\hbox{E}\kern-.125emX}}
\usepackage{balance}

\IEEEoverridecommandlockouts

\usepackage[latin9]{inputenc}
\usepackage{float}
\usepackage{upgreek}
\usepackage{units}
\usepackage{xcolor}
\usepackage{multirow}
\usepackage{arydshln} 

\usepackage{amssymb}
\usepackage{graphicx}
\usepackage{comment}
\usepackage{caption} 
\usepackage{amsmath,amssymb}

\usepackage{lscape}

\usepackage{epsfig}
\usepackage{epstopdf}
\usepackage{simplemargins}
\usepackage{amsmath}

\usepackage{amsfonts}
\usepackage{bm}\usepackage{url}
\usepackage{booktabs}
\usepackage{multicol}

\usepackage{booktabs}
\usepackage{amsthm}
\usepackage{colortbl} 
\usepackage{xcolor} 

\usepackage{color}
\newcommand{\red}{\textcolor{red}}

\newtheorem{theorem}
{Theorem}
\newtheorem{lemma}{Lemma}
\newtheorem{definition}{Definition}

\newtheorem{corollary}{Corollary}
\newtheorem{remark}{Remark}
\newtheorem{proposition}{Proposition}

\def\miR{{\mathcal R}}
\def\miD{{\mathcal D}}
\def\nCp{M}
\def\map{m}
\def\Emb{{\sf E}}
\def\flD{{\sf f}}
\def\muD{{\upmu}}
\def\flR{{\uppsi}}
\def\muR{{\upnu}}

\def\fLP{{\hat{f}}}
\def\muLP{{\hat{\mu}}}
\def\CostR{\Omega}

\def\latencyR{\lambda}

\def\yLP{{\hat{y}}}
\def\yD{{\sf y}}
\def\yR{{\gamma}}
\def\prob{IA-DAG-DTR}
\def\alg{IDAGO}
\def\E{\mathbb{E}} 
\def\CostLP{\sf C_{LP}^*}
\def\Cmilp{\sf C^*_{MILP}} 
\def\CostIDAGO{\sf C^*_{\text{\alg}}}

\def\ratio{CAR}
\def\nwuse{CRF}
\def\burfact{B^k_{uv}}

\begin{document}

\title{Approximation Algorithms for the End-to-End Orchestration of NextG Media Services over the Distributed Compute Continuum}
\title{Information-Aware End-to-End Service Orchestration over the Distributed Compute Continuum}
\title{End-to-End Orchestration of NextG Media Services over the Distributed Compute Continuum}

\author{
Alessandro Mauro 
\IEEEmembership{Student Member, IEEE},
Antonia M. Tulino 
\IEEEmembership{Fellow, IEEE}, 
Jaime Llorca \IEEEmembership{Senior Member, IEEE}
\thanks{A. Mauro, A. Tulino, and J. Llorca are with the DIETI Department, University of Naples Federico II, Italy. Email: {antoniamaria.tulino, alessandro.mauro3, jaime.llorca}@unina.it}
\thanks{A. Tulino and J. Llorca are also with the ECE Department of New York University, Brooklyn, NY. Email: {atulino, jllorca}@nyu.edu }
}

\maketitle

\begin{abstract}
NextG (5G and beyond) networks, through the increasing integration of cloud/edge computing technologies, are becoming highly distributed compute platforms ideally suited to host 
emerging resource-intensive and latency-sensitive applications (e.g., industrial automation, extended reality, distributed AI). The end-to-end orchestration of such demanding applications, which involves function/data placement, flow routing, and joint communication/computation/storage resource allocation, 
requires new models and algorithms able to capture: (i) their disaggregated microservice-based architecture, 
(ii) their complex processing graph structures, including multiple-input multiple-output processing 
stages, and (iii) the opportunities for efficiently sharing and replicating 
data streams that may be useful for multiple functions and/or end users. To this end, we first identify the technical gaps in existing literature that prevent efficiently addressing the optimal orchestration of emerging applications described by information-aware directed acyclic graphs (DAGs). We then leverage the recently proposed Cloud Network Flow optimization framework and a novel functionally-equivalent DAG-to-Forest graph transformation procedure 
to design \alg\, (Information-Aware DAG Orchestration), a polynomial-time 
multi-criteria approximation algorithm for the optimal orchestration of NextG media services over NextG compute-integrated networks.

\end{abstract}


\pagestyle{headings} \pagenumbering{arabic} \setcounter{page}{1}
\thispagestyle{empty}

\section{Introduction}


\IEEEPARstart{N}{ext} generation (NextG) networks (i.e., 5G and beyond), through accelerated efforts in softwarization, programmability, and edge/cloud computing integration, are rapidly evolving toward tightly integrated computation-communication systems that go beyond (i) computation-centric data center networks hosting most of today's applications 
and (ii) communication-centric mobile networks connecting mobile users to cloud-hosted applications. 
We envision NextG cloud-integrated networks becoming highly distributed general-purpose compute platforms, ideally suited to host emerging resource-intensive and latency-sensitive applications, ranging from the automation of physical systems (e.g., smart factories, cities, ports, supply chains) to the augmentation of human experiences (augmented/virtual/extended reality, immersive video, metaverse)~\cite{cai2022compute,sun2019communications,elbamby2018toward}.


\begin{figure}
\centering
\includegraphics[width=1\columnwidth]{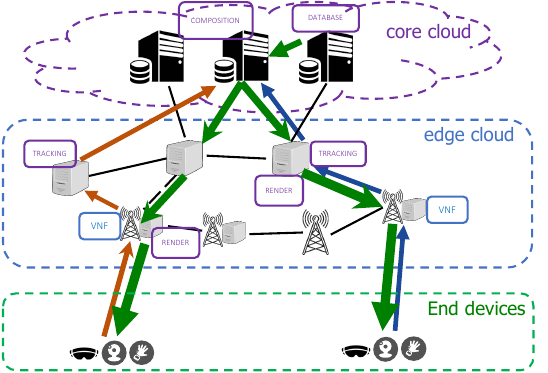}
\caption{\small{Illustration of a microservice-based XR application running over the device-edge-cloud continuum.}}
\label{fig:XR}
\end{figure}


As illustrated in Fig.~\ref{fig:XR} in the context of an extended reality (XR) application, 
key properties of NextG services include:
\begin{itemize}
    \item {\em Disaggregated microservice-based architecture:} NextG services are composed of multiple service functions, typically implemented as containerized microservices.
    \item {\em Multi-stage processing:} NextG services involve multiple processing steps that can be typically classified into three main stages, e.g., tracking/understanding, analysis/synthesis, and rendering/personalization~\cite{cai2022compute}. 
    \item {\em Live streaming and on-demand data pipelines:} NextG services may involve the combined processing of live data streams and pre-stored digital content.
    \item {\em Data/stream sharing/multicast:} NextG services can leverage data caching and stream multicast to efficiently replicate data and live streams useful for multiple service functions and/or end users.
    \item {\em Directed acyclic graph (DAG) structures:} above features and in general, the capability for NextG service functions to process multiple input data streams and generate output streams useful for multiple functions and/or end users, leads to the representation of NextG services via DAGs with arbitrary branching and merging points.
\end{itemize}

\subsection{End-to-end service orchestration}

In this context, NextG network orchestrators will be tasked with the challenging job of optimizing the end-to-end configuration of increasingly complex NextG services over an expanding distributed compute infrastructure. Such an orchestration problem involves three critical decisions:
\begin{itemize}
    \item {\em Function/data placement:} determine the network locations to instantiate and execute each service processing and/or caching function. 
    \item {\em Flow routing}: determine the route each data stream should follow from its producer function location to its consumer function location.
    \item {\em Comm./comp./storage resource allocation:} determine the amount of communication, computation, and storage resources to allocate across the distributed compute continuum in order to support the execution of the service functions and the delivery of associated data streams. 
\end{itemize}

The above decisions should be jointly determined in order to guarantee quality of service (QoS) requirements (i.e., sustaining service rates and guaranteeing end-to-end latency constraints) while minimizing the overall use of the shared physical infrastructure.

While existing solutions for service placement based on virtual network embedding (VNE) formulations \cite{fischer2013virtual,chowdhury2011vineyard,addis2015virtual,pei2019efficiently,agarwal2019vnf,rost2019virtual,rost2019parametrized} may seem natural candidates for optimizing the deployment of NextG services over NextG networks, 
the unique properties of these services, such as the fact that live data streams and pre-stored content can be shared by multiple functions and end users,  
critically break the suitability of VNE models and algorithms. 
In particular, the possibility to replicate data streams, content, and service functions as needed in order to optimize service deployment, breaks the isomorphic (one-to-one) nature of the mapping between a service graph and its instantiation on the physical infrastructure that is a core assumption of VNE-based approaches. Such a limitation comes from the fact that VNE approaches treat services as virtual networks carrying point-to-point traffic demands, {\em without actual knowledge of the information they carry}~\cite{michael2019approximation,Llorca2024CNF}.

\subsection{Contributions}

To this end, in this paper, we leverage the recently proposed Cloud Network Flow (CNFlow) optimization framework that allows formulating the joint placement, routing, and resource allocation problem as an {\em information flow problem over a cloud-augmented network graph}~\cite{Barcelo2015csdp,Barcelo2016jsac,Llorca2024CNF}.

The key benefit of the proposed CNFlow-based formulation lies in its ability to inherently capture the unique nature of NextG service flows, such as:
\begin{itemize}
\item {\em Flow splitting}: streams in NextG services can be splittable or unsplittable; while in some cases, streams can be split into multiple sub-streams to increase resource efficiency, 
some data streams may be required to travel and get processed without splitting (e.g., for video analytics).
\item {\em Flow/function chaining}: streams in NextG services must be chained and processed through the appropriate sequence of service functions according to their input-output relationship established by their corresponding service graph. 
\item {\em Flow scaling}: streams in NextG services can change size as they get processed; streams can either expand (e.g., via decoding/decompression) or shrink (e.g., via video tracking, detection, or compression functions).
\item {\em Flow/function replication}: the fact that streams in NextG services can be shared by multiple functions and/or end users at different locations requires the capability to replicate streams within the network, which in turn leads to the replication of service functions. Flow/function replication is, in essence, a consequence of the {\em multicast} nature of real-time streams in NextG services.
\end{itemize}


While CNFlow allows 
capturing the unique properties of NextG services and networks, the resulting optimization problem, in its {\em unsplitabble} flow version, is still NP-Hard~\cite{Barcelo2016jsac,Poularakis2020Mobihoc,Llorca2024CNF}.

Motivated by the fact that many NextG services require media streams to (i) travel and get processed without splitting (e.g., for video analytics) and (ii) be shared by multiple functions or end users, we seek the design of efficient polynomial-time solutions for the class of unsplittable multicast CNFlow problems that encompass the majority of NextG service orchestration problems. 
To this end, 
we design \alg, the first (to the best of our knowledge) multi-criteria approximation algorithm for the end-to-end distribution (placement, routing, resource allocation) of generic information-aware DAG services over the distributed compute continuum. \prob's main innovation comes from the use of a novel functionally-equivalent DAG-to-Forest graph transformation procedure that allows (i) maximizing flow/function replication opportunities, and (ii) 
adapting existing techniques for information-unaware service tree embeddings to compute information-aware DAG embeddings in polynomial time.

Our contributions are summarized as follows:
\begin{itemize}
\item We provide a CNFlow-based formulation for the optimal distribution (function/data placement, flow routing, resource allocation) of NextG information-aware DAG services over NextG cloud-integrated networks, referred to as \prob, that captures arbitrary flow splitting, chaining, scaling, and replication, and guarantees the support of requested service rates and end-to-end latency requirements while minimizing overall communication, computation, and storage resource costs.
\item We 
design \alg\, (Information-Aware DAG Orchestration), the first multi-criteria approximation algorithm for this class of CNFlow problems. 
\alg\, leverages a novel {\em information-aware DAG-to-Forest service graph transformation} procedure that allows (i) maximizing replication opportunities and (i) adapting LP relaxation, decomposition, and rounding techniques to yield polynomial-time solutions with constant factor multi-criteria approximation guarantees. 
\item We provide extensive simulation results illustrating the performance of \alg\,  in the context of NextG media services, demonstrating fold efficiency improvements when compared to even the optimal (exponential-time) solution to the state-of-the-art information-unaware VNE-based solution. 
\end{itemize}

\section{Related Work and Technical Gaps}
\label{sec:rwork}

\subsection{Service placement without routing}

Max-profit and min-cost versions of the (monolithic) service placement and request assignment problem, where the goal is to place a set of single-function services or applications over a distributed computing network (e.g., mobile edge computing network) and assign (user) requests to placed service instances, has been studied under different resource (communication, computation, storage) constraints~\cite{poularakis2019joint,poularakis2020service}. The work in~\cite{poularakis2020service} provides a taxonomy of existing works addressing the problem under different combinations of communication, computation, and storage/caching constraints, with their own work showing the NP-Hardness and providing a multi-criteria approximation algorithm for the more general version of the problem that includes all three resource-type constraints.  


\subsection{Service placement and routing via VNE}

\begin{figure}
\centering 
\includegraphics[width=0.9\linewidth]{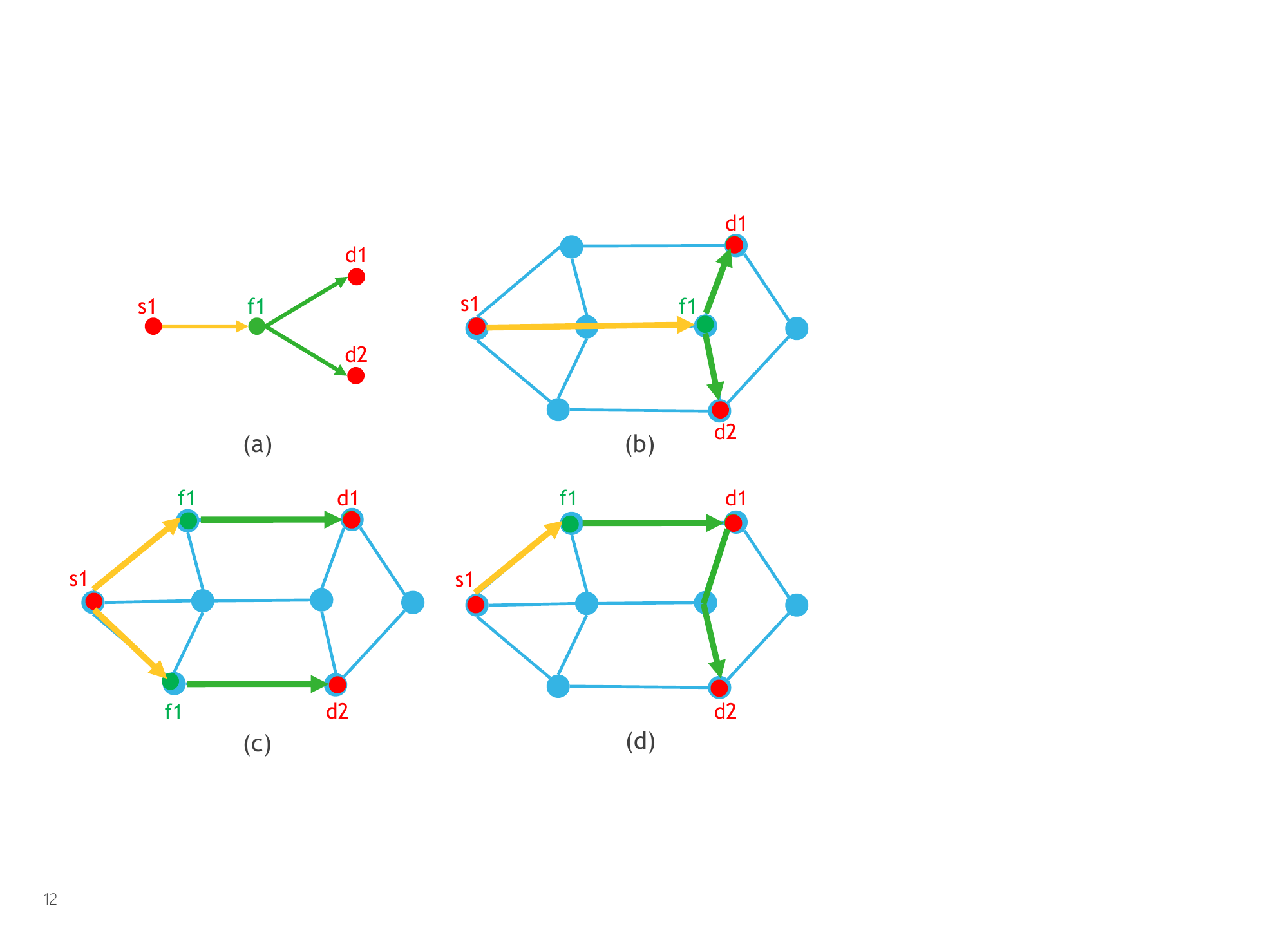}
\vspace{-0.1cm}
\caption{\small{Illustration of the lack of isomorphism between an information-aware service graph and its instantiation into the physical infrastructure: (a) depicts an information-aware service graph with colors indicating the information carried by each data stream; (b), (c), and (d) illustrate three different possible information-aware embedding into an 8-node network.}} 
\vspace{-0.25cm}
\label{fig:multicast}
\end{figure}

The extension to service graph (e.g., service chains, DAGs) placement and routing, where now the goal is to place multiple service functions, represented as vertices of a service graph, and route traffic flows among corresponding functions, was initially studied 
resorting to VNE-type formulations, where the service graph is treated as a virtual network that needs to be embedded into the infrastructure network~\cite{fischer2013virtual,chowdhury2011vineyard,addis2015virtual,pei2019efficiently,agarwal2019vnf,rost2019virtual,rost2019parametrized}. 
A critical aspect of VNE formulations is the isomorphic (or one-to-one) nature of the mapping between the service graph and its instantiation on the physical infrastructure. That is, in VNE, each service function (a vertex in the service graph) must be mapped to exactly one network node (a vertex in the network graph); and each service data stream (an edge in the service graph) to exactly one network path (set of edges in the network graph). 
While such a model is suitable for a number of services and use cases such as services carrying unicast traffic and/or information-unaware data streams, it 
prevents optimizing the replication of data streams and associated functions, which is essential for the optimal deployment of NextG services that include shareable real-time data streams. 
This is illustrated in Fig.~\ref{fig:multicast} in the context of a simple service graph where a source data stream must go through function $f_1$, whose output is requested for consumption by two destination functions. Fig.~\ref{fig:multicast}b  indicates a possible instantiation of the service in an eight-node network, where $f_1$ gets placed at a single location. Fig.~\ref{fig:multicast}c then shows an alternative solution, where $f_1$ gets replicated at two locations, each providing the output stream to be consumed by $d_1$ and $d_2$, respectively. Fig.~\ref{fig:multicast}d shows yet another possible solution where the output of $f_1$ after being delivered to $d_1$ is reused to satisfy the demand of $d_2$. Clearly, non-isomorphic solutions in Fig.~\ref{fig:multicast}c and Fig.~\ref{fig:multicast}d cannot be captured by VNE-based models and formulations. 

\subsection{Service placement and routing via cloud network flow}

To this end, the study of network flow based formulations for the service graph placement and routing problem was initiated 
in \cite{icc15, iot_jsac16}. 
The authors developed network information flow (NIF) based formulations that allow capturing flow chaining, scaling, splitting, and replication, hence accounting for the heterogeneous processing and mixed-cast (unicast and multicast) nature of NextG service flows. 
The resulting model was further extended and generalized under the term Cloud Network Flow (CNFlow) in \cite{Llorca2024CNF}. CNFlow is shown to allow computing the end-to-end orchestration (function/data placement, flow routing, and resource allocation) by solving a single information flow problem on a what is referred to as a {\em cloud-augmented graph} that includes links representing communication, computation, and storage resources. 
In \cite{Llorca2024CNF}, a complexity classification of CNFlow is provided as a function of the splittable vs unsplittable nature as well as of the unicast vs mixed-cast nature of service flows. 
It is shown that the splittable unicast version is Polynomial-time solvable, with fully polynomial-time approximation schemes (FPTAS) given in \cite{info17}. The splittable mixed-cast version is shown to be P-time  under a policy space that admits intra-file network coding~\cite{Llorca2024CNF}. 
Finally, in the context of the unsplittable class, which is known to be NP-Hard, we shall also differentiate between unicast and mixed-cast versions. For the unicast version, shown to be equivalent to VNE~\cite{Llorca2024CNF}, multi-criteria approximation algorithms for tree service graphs were provided in \cite{rost2019virtual, Poularakis2020Mobihoc} and XP-approximations for generic graphs in~\cite{rost2019parametrized}, leaving the unsplittable mixed-cast version as the most challenging class with no known approximation guarantees.



In this paper, we seek the design of the first polynomial-time approximation algorithm for the unsplittable mixed-cast service orchestration problem. 
As illustrated in \cite{Llorca2024CNF}, in the CNFlow framework, mixed-cast computation services or services with shareable flows can be fully characterized via information-aware DAGs, where functions (vertices of the DAG) with multiple outgoing edges represent the need for multiple copies of the same output data stream to be used as input to multiple other functions. 
In addition, 
update-aware data caching can also be captured via information-aware service DAGs, adding to the prominent relevance of this type of service orchestration problems. 
We note that restricted versions of this problem, where replication of data and/or data streams is taken into account only in the presence of multiple destination functions (user-driven multicast), or only for data caching, were addressed in \cite{michael2019approximation}, and in \cite{Poularakis2020Mobihoc} and \cite{cai2023joint}, respectively.

\section{System Model}
\label{sec:model}
\subsection{NextG cloud-integrated network model}

We model a NextG cloud-integrated network (cloud-network for short) as a directed graph $\mathcal G=(\mathcal V,\mathcal E)$, where vertices represent cloud-network nodes (e.g., core cloud nodes, edge cloud nodes, compute-enabled base stations, or end devices with embedded computing resources), and edges represent network links between computing locations. 

\begin{figure}
\centering
\includegraphics[width=0.9\columnwidth]{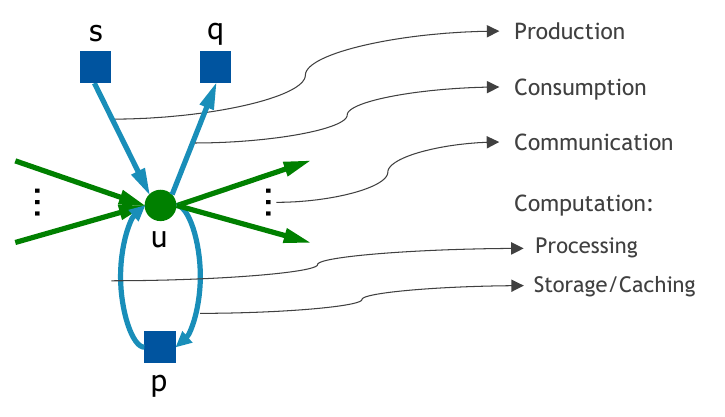}
\caption{\small{Cloud-augmented graph, where green edges represent traditional network links indicating the availability of communication resources for transmitting information between nodes, and blue edges represent production, consumption, and computation capabilities at a given node. 
}}
\label{fig:cloud_aug}
\end{figure}

In line with e.g.,~\cite{Barcelo2015csdp,Barcelo2016jsac,info17}, each node $u\in\cal V$ is further augmented using the gadget in Fig.~\ref{fig:cloud_aug}, where nodes $s$, $q$ and $p$, and 
associated links (shown in blue in Fig.~\ref{fig:cloud_aug}) are used to model the production, consumption, and processing of data streams, respectively. 
The resulting {\em cloud-augmented graph} is denoted by $\mathcal G^a = (\mathcal V^a,\mathcal E^a)$, where $\mathcal V^a = \mathcal V\cup \mathcal V^{p}\cup \mathcal V^{s}\cup \mathcal V^{d}$ and $\mathcal E^a = \mathcal E \cup \mathcal E^{p}\cup \mathcal E^{s}\cup \mathcal E^{d}$, with $\mathcal V^{p}, \mathcal V^{s}, \mathcal V^{d}$ and $\mathcal E^{p}, \mathcal E^{s}, \mathcal E^{d}$ denoting the set of computation, source, and destination nodes, and links, respectively. 

In $\mathcal G^a$, each link $(u,v)\in\mathcal E^a$ is characterized by its capacity $c_{uv}$ and cost $w_{uv}$ parameters. 
In particular, for each communication link $(u,v)\in\mathcal E$, $c_{uv}$ and $w_{uv}$ denote the capacity in communication flow units (e.g., bits per second or bps) 
and the cost per unit flow 
at link $(u,v)\in\mathcal E$, respectively. 
Analogously, for each computation link $(u,v)\in\mathcal E^p$, $c_{uv}$ and $w_{uv}$ denote the capacity in computation flow units (e.g., floating operations per second or FLOPS) 
and the cost per unit flow 
at link $(u,v)\in\mathcal E^p$, respectively. 
In this paper, without loss of generality, we use the set of {\em computation out} links, $\mathcal E^{p+}\subset{\mathcal E^p}$, with origin at a computation node $p\in\mathcal V^p$ and target at a communication node $u\in\mathcal V$, to represent the processing resources (e.g., CPU) available at that computation node/cluster, and the set of {\em computation in} links, $\mathcal E^{p-}\subset{\mathcal E^p}$, with origin at a communication node $u\in\mathcal V$ and target at a computation node $p\in\mathcal V^p$, to represent the memory resources (e.g., RAM) available at that computation node/cluster.\footnote{In Sec.~\ref{sec:extensions}, we describe how to extend the model to capture the allocation of discrete resource blocks such as containers or virtual machines, with predefined processing/memory configurations.}  
Source and destination links $\mathcal E^{s}, \mathcal E^{d}$ are assumed to have zero cost and high enough capacity, acting as network ingress and egress points, respectively.



Finally, we denote by $\mathcal N^{-}(u)$ and $\mathcal N^{+}(u)$ the set of incoming and outgoing links of node $u\in\mathcal V^a$, respectively.

\begin{figure}
\centering 
\includegraphics[width=0.95\columnwidth]{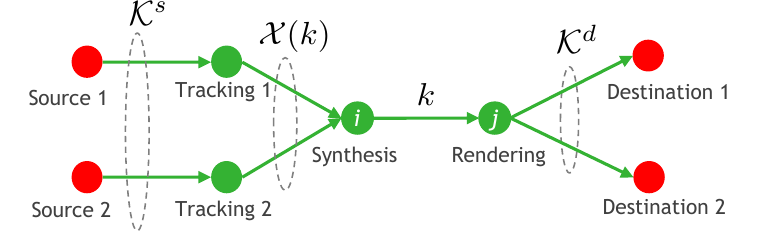}
\caption{\small{Example of a service graph, where edges represent data streams (commodities) and vertices service functions, for a NextG media application, in which streams from two sources go through tracking, synthesis, and rendering functions before being delivered to corresponding destinations.}}
\vspace{-0.25cm}
\label{fig:service_graph}
\end{figure}

\subsection{NextG information-aware service DAG model}

A generic service (or set of services) can be described by a directed acyclic graph (DAG) $\cal R=(\mathcal I, \mathcal K)$, where vertices represent service functions (e.g., stream processing operators) and edges corresponding data streams (or commodities), 
as shown in Fig.~\ref{fig:service_graph}. 



The vertices with no incoming edges of the service graph represent source 
functions that produce source data streams (e.g., video capture), 
and the vertices with no outgoing edges destination 
functions that consume processed data streams (e.g., video display). Source/destination functions may also represent purely ingress/egress points injecting/ejecting data in/out of the network, and are always associated with a {\em fixed and unique location} in the cloud-network (depicted in red in Fig.~\ref{fig:service_graph}), while the remaining functions are subject to placement optimization (depicted in green in Fig.~\ref{fig:service_graph}). 


An edge $k\equiv (i,j)\in\mathcal K$ represents a commodity or data stream produced by function $i\in\mathcal I$ and consumed by function $j\in\mathcal I$. We use $\mathcal X(k)$ to denote the set of incoming edges of node $i\in\mathcal I$, i.e., the set of input commodities required to produce commodity $k\in\mathcal K$ via function $i\in\mathcal I$. 
We denote by $\mathcal K^s\in\mathcal K$ the set of source commodities, i.e., the commodities produced by a source function, and by $\mathcal K^d\in\mathcal K$ the set destination commodities, i.e., the commodities consumed by a destination function. 
We also denote by $s^{\mathcal{K}}(k)\in\mathcal V^s$ the node hosting the 
function producing source commodity $k\in\mathcal K^s$, and by $d^{\mathcal{K}}(k)\in\mathcal V^d$ the node hosting the 
function consuming destination commodity $k\in\mathcal K^d$. 
We also use $\mathcal K^p=\mathcal K\backslash\mathcal K^s$ to denote the set of commodities that are produced by a processing function.\footnote{Note that 
in the special case of communication services, where there are only source and destination functions, then $\mathcal K\equiv\mathcal K^s\equiv\mathcal K^d$ and $\mathcal K^p=\emptyset$.}

Analogously, we define $\mathcal I^s$, $\mathcal I^d$, and $\mathcal I^p$ as the set of source, destination, and computation functions, respectively. We use $s^{\mathcal{I}}(i)\in\mathcal V^s$ to denote the node hosting source function $i \in \mathcal I^s$, and $d^{\mathcal{I}}(i)\in\mathcal V^d$ the node hosting destination function $i \in \mathcal I^d$. Finally, for a given commodity $k\equiv(i,j)$, we denote by  $\mathcal V^{p,\mathcal I}(i)\equiv\mathcal V^{p,\mathcal K}(k)$ the set of computation nodes that can host function $i$ and hence produce commodity $k$.




In $\mathcal R$, each commodity is characterized by its multidimensional rate requirement $R^{k}_{uv}$, which denotes the average rate of commodity $k\in\mathcal K$ when it goes over link $(u,v)\in \mathcal E^{a}$. 
Hence, the rate of a given commodity $k$ will depend on the type of link (resource) $(u,v)$ it goes through. That is, commodity $k$ will impose a certain communication rate (e.g., in bps) when it goes over a communication link $(u,v)\in\mathcal E^c$, a certain processing rate (e.g., in FLOPS) when it goes over a "computation out" link $(u,v)\in\mathcal E^{p+}$, and a certain memory rate (e.g., in bits) when it goes over a "computation in" link $(u,v)\in\mathcal E^{p-}$. 
Note also that communication, processing, and memory rates will be different for different commodities along the service graph, hence capturing the flow scaling nature of NextG services.

Finally, one of the most important aspects of our {\em information-aware} service DAG model, which allows efficiently leveraging the multicast nature of real-time data streams and their possible replication over the network, is the ability to characterize the actual information or content carried by each commodity. As such, we differentiate between the set of commodities $\mathcal K$ and the set of information objects $\mathcal O$, 
and use the surjective {\em information mapping function} $g: \mathcal{K} \to \mathcal{O}$ to indicate the information object $o\in\mathcal O$ associated with each commodity $k\in\mathcal{K}$. 
As shown in the next section, the information mapping function will be key to allow the overlapping of commodity flows that carry the same information, in turn creating opportunities for in-network replication of shareable information flows.


\begin{figure}
    \centering
    \includegraphics[width=1\columnwidth]{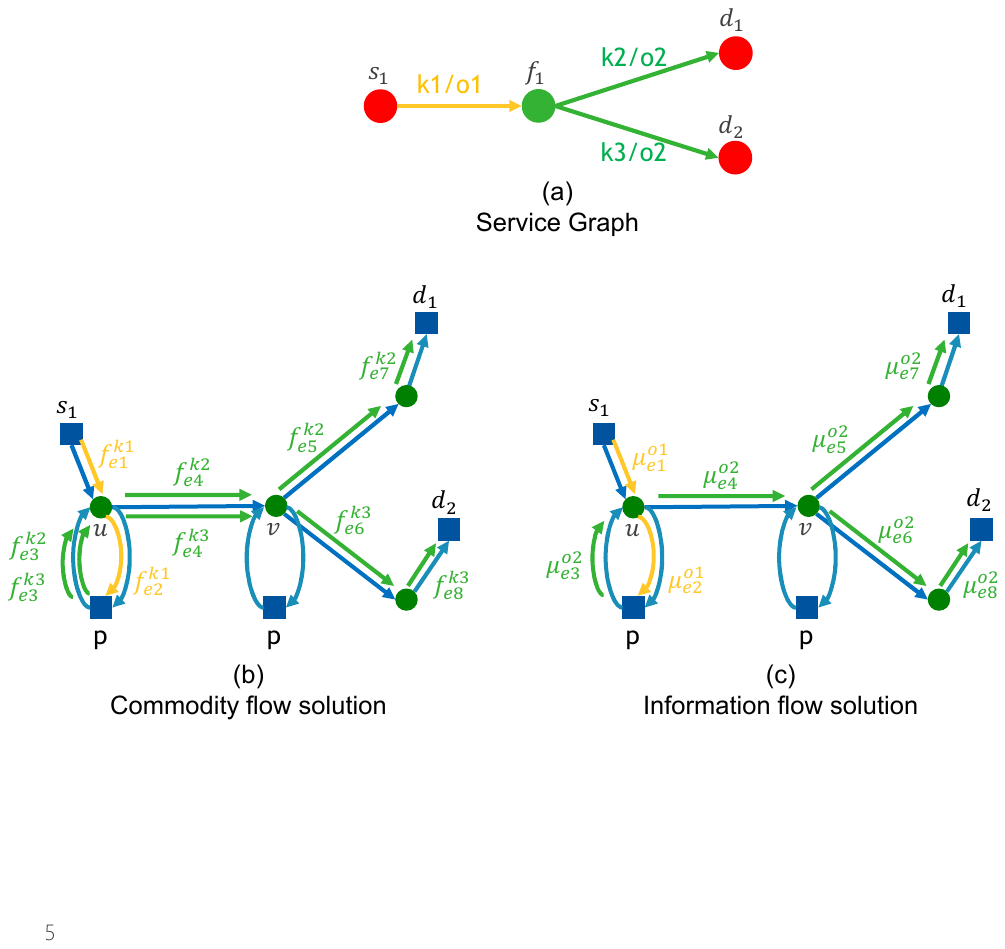}
    \caption{\small{Illustration of how information aware overlapping of commodity flows results in replication of information flows. Commodities $k_2$ and $k_3$ carry the same information and are hence associated with the same object $o_2$. Hence, only one copy of the associated information flow needs to travel over common links $e_3$ and $e_4$.}}
    \label{fig:overlap}
\end{figure}

\section{CNFlow-based Formulation}\label{ProblemFormulation}

We define the information-aware service DAG distribution problem (\prob) as, given a full description of a cloud-network graph $\mathcal G^a$ and a full description of a service graph $\mathcal R$, finding (i) the number of replicas and location of each service function $i\in\mathcal I$, (ii) the routes of each information object $o\in\mathcal O$, and (iii) the allocation of compute, storage, and communication resources, that guarantee given service rates and end-to-end latency constraints, while minimizing overall resource cost.

We provide a CNFlow formulation for the \prob\, problem based on the following variables:

1) {\em Virtual Commodity Flows $\{f_{uv}^{k}\}$:} adimensional binary variables indicating whether commodity $k\in\mathcal{K}$ goes (i.e., is transmitted, processed, or stored) over link $(u,v)\in\mathcal E^a$. 

2) {\em Actual Information Flows $\{\mu^{o}_{uv}\}$ and $\{\mu_{uv}\}$:} real variables indicating the amount of information flow associated with object $o\in\mathcal{O}$ and the total information flow, respectively, going over link $(u,v)\in\mathcal E^a$. 

The resulting mixed integer linear program (MILP) is described as follows:

\begin{subequations}\label{eq:ilp}
\begin{align}
& \text{min}\,\,\displaystyle\sum\limits_{(u,v)\in \mathcal{E}^a} \mu_{uv}w_{uv}  \label{eq:obj}\\
& \text{s.t.}\quad \displaystyle\sum\limits_{v\in \mathcal N^{\!-\!}(u)} f_{vu}^k = \sum_{v\in \mathcal N^{\!+\!}(u)} f_{uv}^k  \qquad\quad \forall u \in \mathcal{V}, k \in \mathcal{K} \label{eq:flowcons}\\
& \qquad  f_{uv}^k =\! \begin{cases}
    f_{vu}^\ell &\!\! \forall k \!\in\! \mathcal{K}^p, \ell \!\in\! \mathcal{X}(k), u \!\in\! \mathcal{V}^p, v\!\in\!\mathcal N^{+\!}(u) \\
    0 & \!\!\text{otherwise}
  \end{cases} \label{eq:chain}\\
&\qquad f_{uv}^k =\! \begin{cases}
   1 & \forall k \!\in\! \mathcal{K}^s, u = s^{\mathcal{K}}(k), v\!\in\!\mathcal N^{+\!}(u) \\
   0 & \text{otherwise}
 \end{cases} \label{eq:source}\\
&\qquad f_{uv}^k =\! \begin{cases}
   1 & \forall k \!\in\! \mathcal{K}^d, v = d^{\mathcal{K}}(k), u\!\in\!\mathcal N^{-\!}(v) \\
   0 & \text{otherwise}
 \end{cases} \label{eq:dest}\\
&\qquad  f_{uv}^k R_{uv}^k \leq \mu_{uv}^o  \qquad \forall (u,v) \in \mathcal{E}^a, k \in \mathcal{K}, o = g(k) \label{eq:mult1}\\
&\qquad \displaystyle\sum\limits_{o \in \mathcal{O}} \mu_{uv}^o \leq \mu_{uv} \leq c_{uv}  \qquad\qquad\qquad\forall (u,v) \in \mathcal{E}^a \label{eq:cap}\\
&\qquad l^{k} = \sum_{(u,v) \in \mathcal{E}^{a}} l^k_{uv} f_{uv}^k \qquad\qquad\qquad\qquad\,\,\,  \forall k \in \mathcal{K}\label{eq:local_latency}\\
&\qquad l_T^k = l^k \qquad\qquad\qquad\qquad\qquad\qquad\quad\,\,\, \forall k \in \mathcal{K}^s \label{eq:cumulative_latency_source}\\
&\qquad l_T^k \geq l^k +l_T^\ell \qquad \qquad\quad\quad \forall k \in \mathcal{K} \backslash \mathcal{K}^{s}, \ell \in \mathcal{X}(k)\label{eq:cumulative_latency_child}\\
&\qquad l_T^k \leq L^k \qquad \qquad \qquad \qquad \qquad \qquad\quad \forall k \in \mathcal{K}^d \label{eq:cumulative_latency_total}\\
&\qquad f_{uv}^k \!\in\! \{0,1\},  \mu_{uv}^o \!\in\!\mathbb R^+, \mu_{uv} \!\in\! \mathbb R^+, l^k \!\in\!\mathbb R^+, l^k_T \!\in\!\mathbb R^+\notag\\
&\qquad\qquad\qquad\qquad\qquad\quad\, \forall (u,v) \in \mathcal{E}^a, k \in \mathcal{K}, o \in \mathcal{O} \label{eq:domain}
\end{align}
\end{subequations}

In \eqref{eq:ilp}, 
the objective is to minimize the total cloud-network resource cost, where recall that edges in $\mathcal E^a$ can represent communication, computation, or storage resources. 

Eq. \eqref{eq:flowcons} states generalized (communication, computation, storage) flow conservation constraints, requiring the total incoming flow to a given communication node $u\in\mathcal V$ for a given commodity $k\in\mathcal K$ to be equal to the total outgoing flow from node $u$ for commodity $k$. 

Eq. \eqref{eq:chain} states flow chaining constraints, which impose that in order to generate commodity $k\in \mathcal{K}$ at the output of computation node $u\in\mathcal V^p$, all input commodities $\ell\in\mathcal X(k)$ must be present at the input of node $u$. 

Eqs. \eqref{eq:source} and \eqref{eq:dest} are source and destination constraints that initialize the ingress/egress of the source/destination commodities at their corresponding source/destination nodes. 

One of the most important elements of this CNFlow formulation is the connection between virtual commodity flows and actual information flows. 
Recall that a unique aspect of NextG services that cannot be captured via VNE models is the sharing of data streams by multiple processing and/or destination functions. Such multicast nature of NextG media streams means that different virtual commodity flows carrying the same information must be able to overlap when going through the same link $(u,v)\in\mathcal E^a$. 
This is assured by Eq. \eqref{eq:mult1}, where we first multiply the commodity flow variables by their corresponding rate requirement and then allow the overlap of the resulting {\em sized} commodity flows that are associated with the same information object.
This is illustrated in Fig.~\ref{fig:overlap}.

The total information flow at a given link $(u,v)\in\mathcal E$ is then computed by summing over all information flows, which is naturally constrained to be no larger than the total capacity of link $(u,v)$, as stated in Eq. \eqref{eq:cap}.

The end-to-end service latency constraints are governed by equations \eqref{eq:local_latency}-\eqref{eq:cumulative_latency_total}. 
Eq. \eqref{eq:local_latency} computes the local latency of commodity $k$, $l^k$, (i.e., the time taken to produce, deliver, and consume a unit of commodity $k$) as the sum, over the links carrying commodity $k$, of the latency to transmit or process a unit of commodity $k$ over the given link, 
denoted by $l^k_{uv}$. 
Eqs. \eqref{eq:cumulative_latency_source}-\eqref{eq:cumulative_latency_child} compute the cumulative latency of commodity $k$, $l^k_T$, which represents the service latency that has been accumulated until the consumption of commodity $k$. Eq. \eqref{eq:cumulative_latency_source} first sets the cumulative latency to be equal to the local latency for all source commodities. Eq. \eqref{eq:cumulative_latency_child} then computes the cumulative latency for all remaining commodities recursively by setting the cumulative latency of commodity $k$ to be larger than or equal to the local latency of commodity $k$ plus the cumulative latency of input commodity $l$, for all input commodities in $\mathcal{X}(k)$. Lastly, Eq. \eqref{eq:cumulative_latency_total} imposes the cumulative latency at each destination commodity to be no greater than the maximum allowed service latency $L^k$.\footnote{Note that this model allows different maximum service latency for each destination commodity.}

Finally, Eq. \eqref{eq:domain} imposes the binary nature of commodity flow variables and the real positive nature of information flow and latency variables.


\begin{remark}
While, for ease of exposition, we start with a continuous flow-based cost model, in section \ref{sec:extensions}, we show how to extend the cost model to account for the allocation of a discrete number of resource blocks.
\end{remark}

\begin{table*}[]
    \centering
    \begin{tabular}{|p{1.5in}|p{5.2in}|}
    \hline
    Notation & Description \\
    \hline
    $\mathcal G = (\mathcal V, \mathcal E)$; $\mathcal G^a = (\mathcal V^a, \mathcal E^a)$ & Network graph, associated nodes ($\mathcal V$) and links ($\mathcal E$); Cloud-augmented graph, associated nodes ($\mathcal V^a$) and links ($\mathcal E^a$).\\

    
    $\mathcal V^p; \mathcal V^s; \mathcal V^d$ & Computation nodes; Source nodes; Destination nodes.\\

    $\mathcal E^c; \mathcal E^p; \mathcal E^s; \mathcal E^d$ & Communication links; Computation links; Source links; Destination links.\\

    $\mathcal E^{p-}; \mathcal E^{p+}$ & Computation in links (storage resources); Computation out links (processing resources).\\

    $\mathcal N^{-}(u); \mathcal N^{+}(u)$ & Incoming and outgoing neighbors of node $u\in\mathcal V^a$.\\

    $c_{uv}; w_{uv}$ & Capacity and cost of link $(u,v)$.\\
    
    $\mathcal R = (\mathcal I, \mathcal K)$ & Information-aware service (or service collection) graph, composed of functions ($\mathcal I$) and commodities ($\mathcal K$).\\
    
    $\mathcal I^s; \mathcal I^d; \mathcal I^p$ & Source functions; Destination functions; Computation functions.\\
    
    $\mathcal K^s; \mathcal K^d; \mathcal K^p$ & Source commodities; Destination commodities; Processing commodities. \\


        
     $\mathcal X(k)$ & Set of input commodities required to produce commodity $k \in \mathcal K$.\\
     
    $s^{\mathcal K}(k); d^{\mathcal K}(k)$ & Source node hosting the function producing commodity $k \in \mathcal K^s$; Destination node hosting the function consuming commodity $k \in \mathcal K^d$.\\
    
        
    $s^{\mathcal I}(i); d^{\mathcal I}(i)$ & Node hosting source function $i \in \mathcal I^s$; Node hosting destination function $i \in \mathcal I^d$.\\


    $\mathcal V^{p,\mathcal I}(i)\equiv\mathcal V^{p,\mathcal K}(k)$ & Computation nodes that can host function $i$ and hence produce commodity $k$.\\

    $R^k_{uv}$ & Rate of commodity $k \in \mathcal K$ when it goes over link $(u,v) \in \mathcal E^a$.\\

    $\mathcal O$; \quad $g:\mathcal K \rightarrow O$ & Set of information objects; Information mapping function.\\

    $f_{uv}^k; \mu_{uv}^o; \mu_{uv}$ & Virtual commodity flow,  actual information object flow, and actual information flow variables of MILP \eqref{eq:ilp}.\\    

    $l^k_{uv}; l^k; l^k_T; L^k$ & Latency to transmit or process a unit of commodity $k$ over link $(u,v)$; Local latency of commodity $k$; Cumulative latency of commodity $k$; Maximum service latency associated with destination commodity $k$.\\

    $c^T_{uv}; c^b_{uv}; w^b_{uv}; y_{uv}; \burfact$ & Total number of blocks; Capacity per block; Cost per block; Allocated blocks; Burstiness factor.\\

    \hline
    \end{tabular}
    \caption{\small{Main system model notation.}}
    \label{tab:notations_model}
\end{table*}

\section{The \alg\, Algorithm} 
\label{sec:approximation}

Recall that the unsplittable flow nature of \prob\, already renders the problem NP-Hard. In fact, the information-unaware unsplittable DAG orchestration problem can be reduced to the VNE problem, which is already NP-Hard. 

In the following, we introduce \alg, 
to the best of our knowledge, the first polynomial-time multi-criteria approximation algorithm for \prob.

The inspiration behind \alg's design is driven by the following key observations:
\begin{itemize}
    \item Polynomial-time 
    multi-criteria approximation algorithms exist for the unsplittable information-unaware version of \prob\, (reducible to VNE), for service trees \cite{rost2019virtual}.
    \item Information-awareness can be efficiently captured by the CNFlow formulation \eqref{eq:ilp} without breaking the linearity nor increasing the complexity of the information-unaware VNE-based formulation.
    \item The splittable version of  \prob\, is polynomial-time solvable, e.g., via the linear programming relaxation of MILP \eqref{eq:ilp}.
\end{itemize}


Given these observations, \alg\, is designed based on the following key steps:
\begin{enumerate}
\item Information-aware DAG-to-Forest {\bf service graph transformation} procedure that obtains a functionally equivalent transformation of the original service DAG into a set of tree graphs.
\item {\bf LP relaxation} of information-aware MILP \eqref{eq:ilp} for the transformed service graph (set of trees).
\item {\bf Decomposition} of LP commodity flow solution associated with each service tree into a convex combination of valid embeddings.
\item {\bf Randomized rounding} procedure to select an embedding for each service tree. 
\item Composition of overall service forest embedding and computation of {\bf information flow} solution.  
\end{enumerate}

\begin{figure}
    \centering
    \includegraphics[width=1\linewidth]{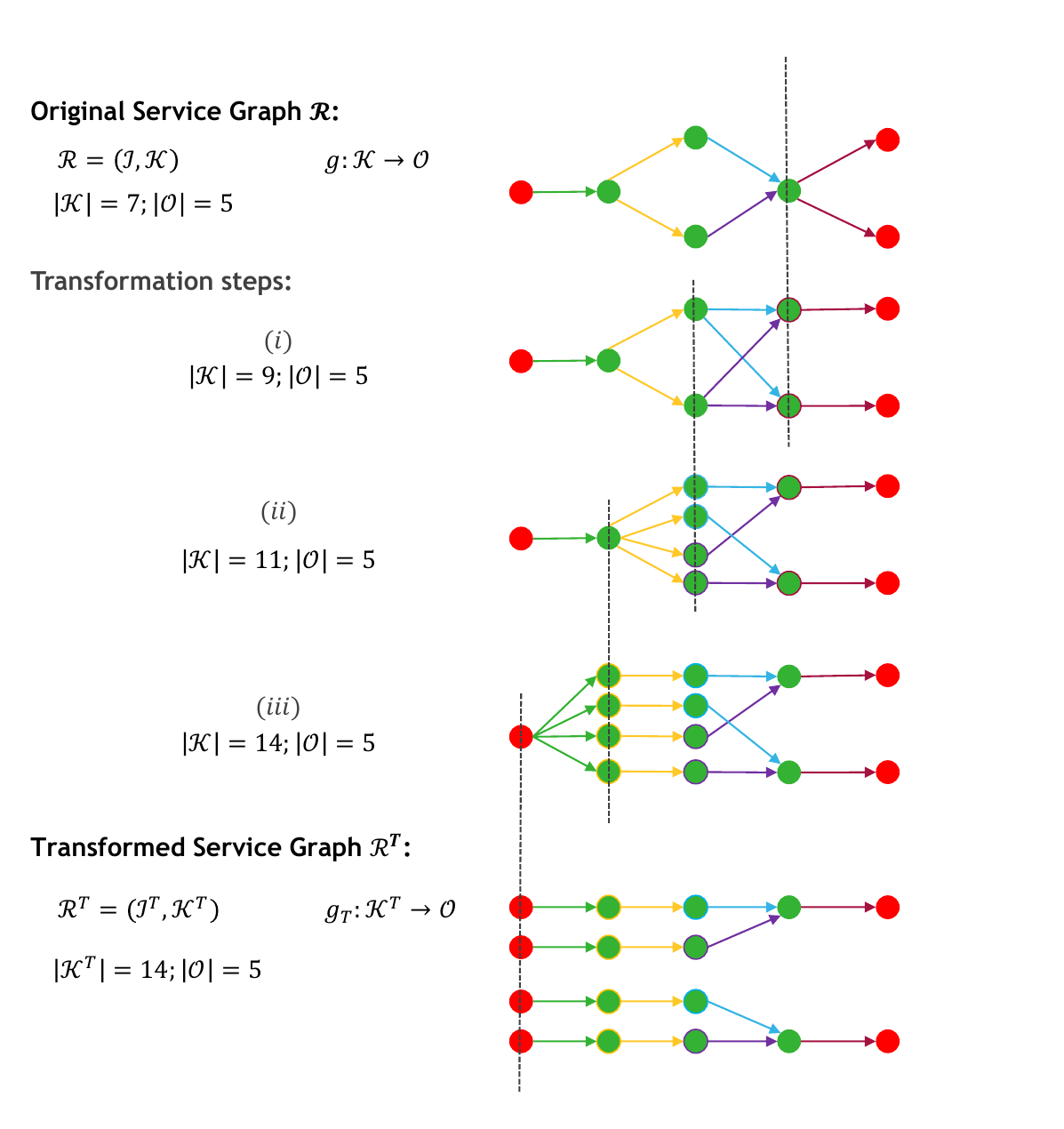}
    \caption{\small{DAG-to-Forest service graph transformation process. In the figure, commodities associated with the same information object are represented with the same edge color.}}
    \label{fig:dagtransformation}
\end{figure}

\subsection{DAG-to-Forest service graph transformation}\label{sec:dagtransform}


One of the key novelties of our approach is an information-aware DAG-to-Forest graph transformation procedure that transforms the original service DAG into a functionally-equivalent set of trees, i.e., a forest, that not only (i) facilitates the adaptation of approximation algorithms for tree graphs, but also (ii) maximizes flow/function replication opportunities. 

The transformation process 
works by systematically eliminating branching points in the service DAG, 
replicating each function that produces more than one output commodity, creating one replica for each output commodity. The end result is a set of trees, each rooted at each destination function in $\mathcal I^d$, as shown in Fig.~\ref{fig:dagtransformation}.

The transformation process traverses the graph backward from the destination functions toward the source functions. 
Since the destination functions do not have outgoing edges, the process starts at the set of functions one hop away from the destinations. The process then goes to the functions two hops away from the destinations, and continues until reaching the source functions. At each stage, the transformation replicates each function that produces multiple commodities creating one replica for each output commodity. Importantly, the set of input commodities is also replicated for each function replica, with each replicated commodity maintaining its original attributes, including its rate requirements and, critically, its associated information object, as illustrated in Fig.~\ref{fig:dagtransformation}. 


Applying the DAG-to-Forest transformation procedure to a given service DAG $\mathcal R = (\mathcal I, \mathcal K)$, results in a forest graph  $\mathcal R^T=(\mathcal I^T, \mathcal K^T)$ containing one tree for each destination function.\footnote{While, in general, each tree $\mathcal R^{T,\phi}$ may contain multiple destination commodities, it can only contain a single destination function, represented by the root of the tree, i.e., the only vertex with no outgoing edges.} 
Importantly, the transformed service forest $\mathcal R^T$ is now associated with a new information mapping function $g_T:\mathcal K^T\rightarrow \mathcal O$ that maps a new set of commodities $\mathcal K^T$ (with $|\mathcal K^T| >= |\mathcal K|$) to the same set of information objects $\mathcal O$.



\begin{figure}[h]
    \centering
    \includegraphics[width=\linewidth]{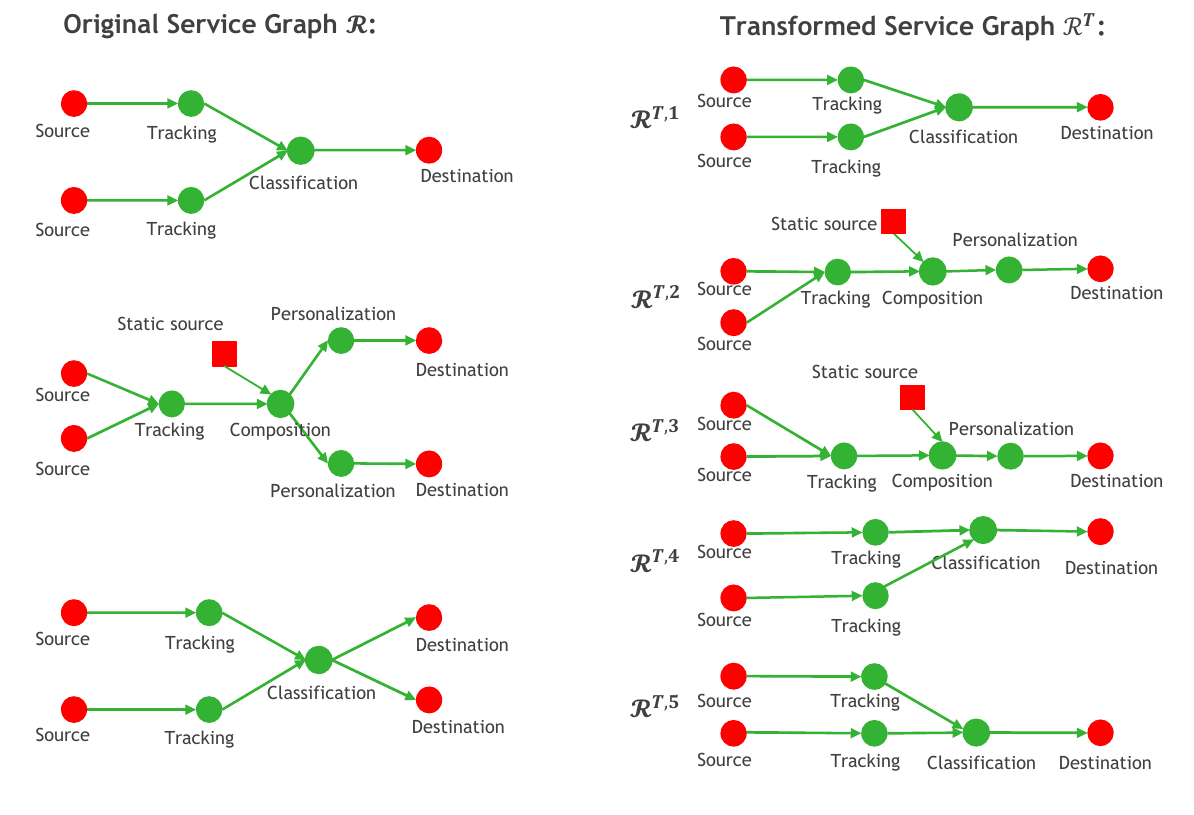}
    \caption{\small{Example of a Service $\mathcal R$ composed by 3 connected components and the resulting DAG-to-Forest transformation composed of one tree for each destination function.}}
    \label{fig:multiple_component}
\end{figure}

\begin{remark}
Recall that $\mathcal R$ may represent a set of services, identified as multiple connected components, as illustrated in Fig.~\ref{fig:multiple_component}. Furthermore, during the DAG-to-Forest transformation procedure, each component is either already a tree or gets decomposed into a set of trees. 
We use $\mathcal R^{T,\phi}$ to denote the $\phi-th$ connected component (tree) of the transformed service graph (forest) $\mathcal R^T$.
\end{remark}

\begin{remark}
Note that the DAG-to-Forest graph transformation, by only replicating functions with multiple same-information output commodities, as well as 
their corresponding input commodities and associated attributes, guarantees the functional equivalence between the original service graph $\mathcal R$ and the resulting transformed graph $\mathcal R^T$. In fact, it is immediate to show that an embedding of an information-aware DAG can always be obtained via an embedding of its transformed graph, by collocating resulting function replicas and associated commodities.
Recall that while the number of commodities in $\mathcal R^T$ may be larger than those in $\mathcal R$, the set of information objects is always identical. 
\end{remark}

\begin{remark}
While it is clear that a given commodity $k\in\mathcal K^T$, indeed an edge graph $\mathcal R^T$, can only belong to one connected component (or tree), note that there may be multiple commodities in different connected components associated with the same information object $o\in\mathcal O$.
\end{remark}

We end this subsection by stressing that the tree structure of the transformed service graph will be key to enabling the design of a multi-criteria approximation algorithm for the optimal orchestration of information-aware DAGs that (i) leverage existing VNE-based approximation algorithms designed for information-unaware trees, while (ii) still exploiting information-awareness to maximize replication opportunities and reduce overall orchestration cost, as shown in the following sections.



\subsection{\alg\, algorithm description}
\label{sebsectionIDAGO}
Before describing the \alg\, algorithm, we shall introduce the following definition:

\begin{definition}
\label{validmapping}
A valid embedding of service graph $\mathcal R$ to cloud-network graph $\mathcal G^a$ is a pair of mappings $\Emb=(\map^{\mathcal I},\map^{\mathcal K})$, i.e., a function mapping $\map^{\mathcal I}$ and a commodity mapping $\map^{\mathcal K}$, where:
\begin{itemize}
    \item The function mapping $\map^{\mathcal I}: \mathcal I\rightarrow \mathcal V^a$, maps each source function $i\in\mathcal I^s$ to its fixed location $s^{\mathcal I}(i)$, each destination function $i\in\mathcal I^d$ to its fixed location $d^{\mathcal I}(i)$, and each processing function $i\in\mathcal I^p$ to a valid computation node, i.e., a node in $\mathcal V^{p,\mathcal I}(i)$. 
    \item The commodity mapping $\map^{\mathcal K}: \mathcal K\rightarrow \mathcal E^a$, 
    maps each commodity $k=(i,j)\in\mathcal K$ to a valid path in $\mathcal G^a$, i.e., a path starting at node $m^{\mathcal I}(i)$ and ending at node $m^{\mathcal I}(j)$. 
\end{itemize}
\end{definition}


Given the system model defined in Sec.~\ref{sec:model} and Definition~\ref{validmapping}, the \alg\, algorithm proceeds as follows:

{\bf \alg\, Algorithm }
\hspace{1cm}
{\em 
\begin{enumerate}
\item {\bf Step 1}: Transform service graph collection $\mathcal R$ via the DAG-to-Forest transformation procedure, obtaining 
$$\mathcal{R}^T = \bigcup_{\phi=1}^\nCp \mathcal{R}^{T,\phi},$$
with $\nCp = |\mathcal K^d|$ denoting the total number of connected components or service trees in $\mathcal{R}^T$. 
\item {\bf Step 2}: Solve LP relaxation of MILP \eqref{eq:ilp} for $\mathcal R^T$ and denote by $\{\fLP^k_{uv}\}$ the associated commodity flow solution. 

\item {\bf Step 3}: 
For each service tree $\mathcal{R}^{T,\phi} \in \mathcal{R}^T$, decompose the associated LP commodity flow solution into a convex combination of valid embeddings using \textbf{Algorithm~\ref{disjoint}}. 

A decomposition $\mathcal D^{\phi}$ of service component $\miR^{T,\phi}$ consists of a set of valid embeddings and associated probability pairs: 
$$\miD^{\phi}= \left  \{ (\Emb^{\phi}_{1}, p^\phi_1) \ldots  (\Emb^{\phi}_{N_\phi}, p^\phi_{N_\phi}) \right \} $$
with $N_{\phi}$ the number of embeddings associated with service tree $\phi$, and where each valid embedding $\Emb^{\phi}_n$ 
is composed of a function mapping 
and a commodity mapping, 
$\Emb^\phi_n=(\map^{\mathcal I, \phi}_{n},
\map^{\mathcal K, \phi}_{n})$,
and a probability value $0<p^\phi_n \leq 1$ with $\sum_{n=1}^{N_\phi} p_n^\phi =1$ (see Lemma \ref{decomp}).

\item {\bf Step 4}: For each service tree $\mathcal{R}^{T,\phi} \in \mathcal{R}^T$: 
\begin{itemize}
    \item Draw embedding $\Emb^{\phi}$ from set $\{ \Emb^{\phi}_{1}, \ldots, \Emb^{\phi}_{N_\phi} \}$ according to probability distribution
    $[ p^{\phi}_{1} \ldots p^{\phi}_{N_\phi} ]$.
    \item For all $(u,v) \! \in  \! \mathcal E^a$ and $k \! \in  \! \mathcal K^{T,\phi} $, compute commodity flow as:\footnote{For clarity, we use sans serif to indicate flow variables computed by the \alg\, algorithm.} 
\begin{equation*}
\flD^k_{uv}(\Emb^{\phi}) =   
    \begin{cases}
        1 \quad  \text{if } (u,v) \in  \map^{\mathcal K, \phi} (k) \\
        0 \quad  \text{otherwise}
    \end{cases}
\end{equation*}
\end{itemize}

\item \textbf{Step 5}: Let $\underline{\Emb}=[\Emb^{1}, \Emb^{2}, \ldots, \Emb^{\nCp}]$
denote the chosen embedding for the service collection $\mathcal R^T$. 
Then:
\begin{itemize}
    \item For all $(u,v) \! \in  \! \mathcal E^a$ and $k \! \in  \! \mathcal K^{T}$, set the service collection commodity flow as: 
 \begin{equation*}
\flD^k_{uv}(\underline{\Emb}) =  
\flD^k_{uv}(\Emb^\phi) \text{  with  } \phi  \text{ such that} \,  k \! \in  \! \mathcal K^{T,\phi}
\end{equation*}


\item Compute information object flow as:
\begin{equation*}
    \muD^{o}_{uv}(\underline{\Emb}) \! =\!\! \!\max_{k\in g_T^{-1}(o)} \!\! \! \left\{ R^k_{uv} \flD^{k}_{uv}(\underline{\Emb})
    \right \}, \forall (u,v)\in\mathcal E^a, o\in\mathcal{O}
\end{equation*}
   \item Compute  service collection 
information flow as:
\begin{equation*}
    \muD_{uv}(\underline{\Emb}) = \sum_{o\in \mathcal O} \muD^o_{uv}(\underline{\Emb}) \qquad \forall (u,v)\in\mathcal E^a
\end{equation*}

\end{itemize}

\item  \textbf{Step 6}: Repeat {Step 4}  and {Step 5} until the solution satisfies the desired accuracy or the maximal rounding tries are exceeded.
\end{enumerate}
}

\begin{figure}[H]
    \centering
    \includegraphics[width=0.8\linewidth]{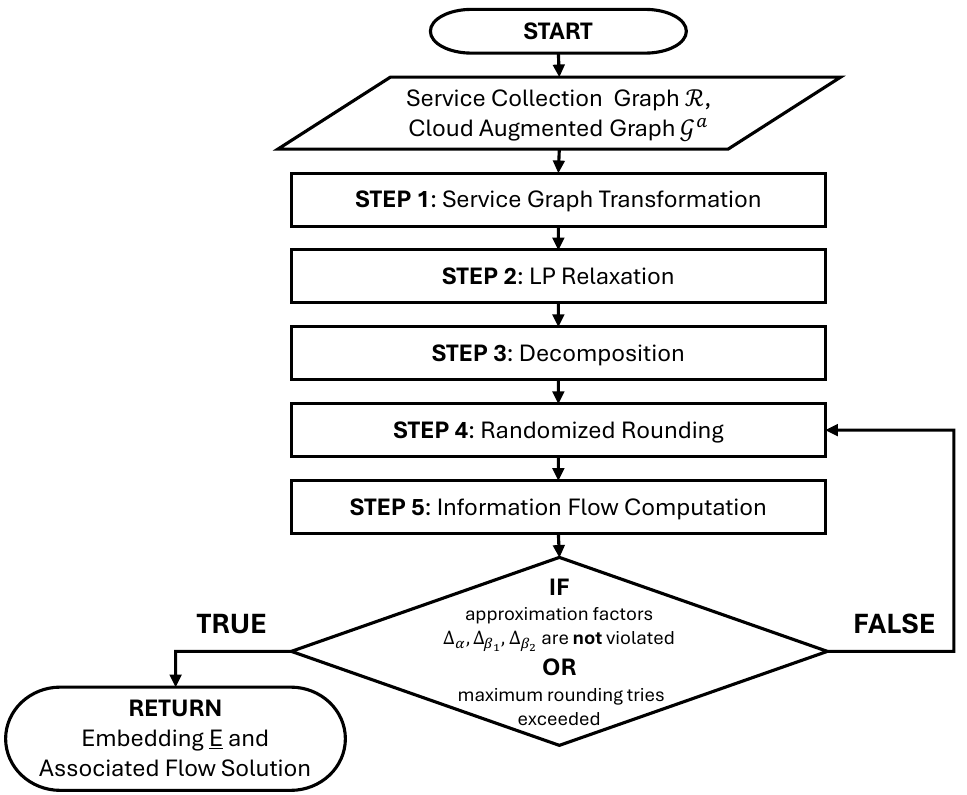}
    \vspace{0.2cm}
    \caption{\small{\alg\, Algorithm flowchart.}} 
    \label{fig:dagdtr}
\end{figure}

The flow chart illustrated in Fig.~\ref{fig:dagdtr} provides a visual description of the entire \alg\, algorithm.

\begin{algorithm}[H]
\label{block_digrama_algorith1}
\caption{Decomposition step of \alg}
\begin{algorithmic}[1]  
\renewcommand{\algorithmicrequire}{\textbf{Input:}}  
\renewcommand{\algorithmicensure}{\textbf{Output:}}  
\REQUIRE Service tree $\mathcal R^{T,\phi}=(\mathcal I^{T,\phi},\mathcal K^{T,\phi})$ and 
LP commodity flow solution $\{\fLP^{k}_{uv}, \, \forall k\in\mathcal K^{T,\phi},\forall (u,v)\in\mathcal E^a\}$ 
 	\STATE{$\mathcal D^{\phi} \leftarrow \emptyset$,  $n \leftarrow 1$} 
  \STATE{${\sf dest}\leftarrow d^{\mathcal I}(i),\quad \{ i \} \equiv\mathcal I^{d,T,\phi}$}
  \STATE{$\fLP^\phi_{\rightarrow d}= \sum_{k \in  \mathcal{K}^{d,T,\phi}}\sum_{u\in\mathcal N^{\!-\!}({\sf dest})}\fLP_{u\,{\sf dest}}^k
  $}
 \WHILE {$\fLP^{\phi}_{\rightarrow{d}}> 0$}
		\STATE{$\Emb^\phi_n=(\map_n^{\mathcal I,\phi},\map_n^{\mathcal K,\phi}) \leftarrow (\emptyset,\emptyset)$}
		\STATE{$\mathcal W \leftarrow \emptyset$}
		\STATE{${\mathcal Q} \leftarrow$ $\mathcal{I}^{d,T,\phi}$ }
            \STATE{$\map^{\mathcal I, \phi}_n(i) = {\sf dest}$}
		\WHILE {$|{\mathcal Q}|>0$}
			\STATE{pick a function $j \in {\mathcal Q}$ and set ${\mathcal Q} \leftarrow {\mathcal Q} \backslash \{j\}$}
   
			\FOR{\textbf{each incoming commodity} $k\equiv(i,j)$ } 
			  \STATE{update $\map_n^{\mathcal K,\phi}(k)$ and $\map_n^{\mathcal I,\phi}(i)$ according to} \textbf{Procedure 1}  
               \STATE{${\mathcal Q} \leftarrow {\mathcal Q} \cup\{i\}$}
			\ENDFOR
		\ENDWHILE 
  	\STATE{$\mathcal W \leftarrow   \{\fLP^{k}_{uv}| \,(u,v)\in\map^{\mathcal K, \phi}_n(k), k\in\mathcal K^{T,\phi}\}$} 
		\STATE{$p^\phi_n \leftarrow \min (\mathcal W)$}
        \STATE{$\fLP^k_{uv} \leftarrow \fLP^k_{uv} - p^\phi_n\cdot 1\Big\{(u,v) \in \map_n^{\mathcal K,\phi}(k), k \in \mathcal K^{T,\phi}\Big\}$}
		\STATE{$\mathcal D^{\phi} \leftarrow \mathcal D^{\phi} \cup ( \Emb^\phi_n,p^\phi_n)$ }
            \STATE{$n \leftarrow n+1$}
            
	\ENDWHILE  
\ENSURE  Decomposition $\miD^{\phi}$ 
\end{algorithmic} 
\label{disjoint}
\end{algorithm}

\floatname{algorithm}{Procedure 1}
\label{block_digrama_Procedure1}
\begin{algorithm}[H]
\renewcommand{\thealgorithm}{}
\caption{Commodity Mapping}
\begin{algorithmic}[1]  
\renewcommand{\algorithmicrequire}{\textbf{Input:}}  
\renewcommand{\algorithmicensure}{\textbf{Output:}}  
\REQUIRE Commodity $k\equiv(i,j)$ 
 	\STATE{$\mathcal{P} \leftarrow$ path with non-zero flow of commodity $k$ that starts at a production function (i.e., a computation or a source function) and ends at the location of the already mapped function $j$, i.e., $\map^{\mathcal I, \phi}_{n}(j)$}
    \STATE{$\map_n^{\mathcal K,\phi}(k) \leftarrow \mathcal{P}$}
    \STATE{$\map_n^{\mathcal{I}, \phi}(i) \leftarrow$ start node of $\mathcal{P}$}
\ENSURE $\map_n^{\mathcal K,\phi}(k)$, $\map_n^{\mathcal I,\phi}(i)$
\end{algorithmic} 
\label{procedure1}
\end{algorithm}

\bigskip

\bigskip

\begin{remark}
    At each iteration of Algorithm~\ref{disjoint} (the decomposition step of \alg), 
    the service tree $\mathcal{R}^{T,\phi}$ is traversed from the unique destination function in $\mathcal I^{d,T,\phi}$ toward the source functions in $\mathcal K^{s,T,\phi}$, finding, for each commodity, a valid path with non-zero flow in the cloud-network graph $\mathcal G^a$. At the end of each iteration, the commodity flow variables are reduced by the minimum commodity flow value associated with all mapped commodities. The decomposition process ends when the residual incoming flow to the destination node hosting the unique destination function reaches zero.
\end{remark}

\begin{remark}
After the decompostion of the LP commodity flow solution into a convex combination of valid embeddings for each service tree (Step 4), \alg\, leverages a randomized rounding procedure to probabilistically select a valid embedding for each service tree (Step 5), whose union and associated information flow solution (Step 6) provides approximation guarantees, as analized in Sec.~\ref{sec:analysis}. 
\end{remark}

\subsection{Polynomial Runtime}



Recall that \alg\, is able to leverage existing polynomial-time LP relaxation, decomposition, and rounding techniques for information-unaware tree embeddings~\cite{rost2019virtual} in Steps 2-4, by first going through an information-aware service graph transformation procedure in Step 1, and then bringing back information-awareness to compute actual information flows in Step 5. 
Since Step 1 is an offline step that, in any case, can be executed in polynomial time (using a graph traversal algorithm on a graph whose number of nodes is $O(|\mathcal I| |\mathcal K|$) 
and Step 5 is linear as it just computes every information flow variable via constant-time max operations, \alg \, is able to solve information-aware service orchestration problems maintaining the polynomial-time complexity of existing information-unaware tree embedding algorithms.

\begin{table*}[]
    \centering
    \begin{tabular}{|p{1.7in}|p{5in}|}
    \hline
    Notation & Description \\
    \hline
    $\mathcal R^T = (\mathcal I^T, \mathcal K^T)$ & Transformed service forest and associated functions and commodities. \\ 
    $\mathcal R^{T,\phi} = (\mathcal I^{T,\phi}, \mathcal K^{T,\phi})$ & $\phi-th$ connected component (tree) of transformed service graph (forest) $\mathcal R^T$.\\

    $\mathcal I^{d,T,\phi}; \mathcal K^{s,T,\phi}$ & Set of destination functions of $\mathcal R^{T,\phi}$; Set of source commodities of $\mathcal R^{T,\phi}$.\\

    $g_T:\mathcal K^T \rightarrow O$ & Information mapping function for transformed service graph $\mathcal R^T$.\\

    $M=|\mathcal K^d|$ & Number of connected components (service trees) in transformed service graph $\mathcal R^T$.\\

    $\fLP^k_{uv}; \muLP^o_{uv}; \muLP_{uv}$ & Fractional commodity flow, object flow, and information flow solution from LP relaxation of MILP \eqref{eq:ilp}. \\

    $\fLP^{\phi}_{\rightarrow{d}}$ & Fractional commodity flow consumed by the single destination function of $\mathcal R^{T,\phi}$ throughout the Decomposition step of \alg.\\

    $\Emb^{\phi}_{n} = (\map_n^{\mathcal I,\phi}, m_n^{\mathcal K, \phi})$ & Embedding, composed of function mapping and commodity mapping of service tree $\phi$ computed at iteration $n$ of the Decomposition step of \alg.\\

    $\miD^{\phi}= \left  \{ (\Emb^{\phi}_{1}, p^\phi_1) \ldots  (\Emb^{\phi}_{N_\phi}, p^\phi_{N_\phi}) \right \}$ & Decomposition of a service tree $\phi$, composed of a set of valid embeddings and associated probabilities.\\

    $N_\phi$ & Number of embeddings of service tree $\phi$ computed by \alg.\\

    $\flD^k_{uv}; \muD^o_{uv}; \muD_{uv}$ & Commodity flow, object flow, and information flow variables computed by \alg.\\
    
    
    $\underline{\Emb}=[\Emb^{1}, \Emb^{2}, \ldots, \Emb^{\nCp}]$ & Embedding of service collection $\mathcal R^T$ computed by \alg.\\

    $\ratio; \quad \nwuse$ & Cost Approximation Ratio; Capacity Relaxation Factor.\\

    \hline
    \end{tabular}
    \caption{\small{Main notation related to the \alg\, algorithm.}}
    \label{tab:notations_idago}
\end{table*}
\section{Performance Analysis of \alg}\label{sec:analysis}




\textit{\bf Notation:} 
$\mathbb{E}[X]$ denotes the expected value of the random  variable $X$, and
$1\left\{ \mathcal A \right\}$ the indicator function of event $\mathcal A$. 
Recall that sans serif symbols indicate variables computed by the \alg\, algorithm. Finally, quantities used in this section, not included in Tables~\ref{tab:notations_model} and \ref{tab:notations_idago}, are defined in Table~\ref{tab:notations_extended} of the Supplementary Material.

The objective of this section is to demonstrate that IDAGO provides probabilistic multi-criteria approximation guarantees. 
That is, \alg\, obtains a solution 
that approximates the objective function value of MILP~\eqref{eq:ilp} on $\mathcal R^T$ by a specified bound, while limiting the violation of the constraints by a given factor, with high probability.
Formal guarantees are stated in our main result, Theorem~\ref{thm2}.\footnote{In the following, unless specified, the solution to MILP~\eqref{eq:ilp} is understood to be on transformed graph $\mathcal R^T$.}


Before proceeding with the formal proof, we provide the following guiding remarks to clarify the unique nature and challenges
associated with the analysis of information-aware service graph orchestration algorithms. 

\begin{remark}  
Recall that the graph transformation procedure described in Sec. \ref{sec:dagtransform} converts a DAG service $\mathcal R$ into a functionally-equivalent forest $\mathcal R^T$ composed of (possibly) multiple connected components. Specifically, it create a tree for each destination function $k\in\mathcal K^d$. 
This allows Algorithm \ref{disjoint} to focus on information-unaware tree embeddings, for which efficient solutions exist~\cite{rost2019virtual}. 
Importantly, IDAGO then leverages the fact that the LP solution used 
in Algorithm~\ref{disjoint} satisfies the information-aware overlapping constraints \eqref{eq:mult1} to compute the information flow solution by simply taking the max over the commodity flows that carry the same information object, as described in Step 5. 
As a result, while certain algorithms and analyses proposed in existing literature remain applicable at the commodity flow level, 
new tools are needed to analyze the effect of
information-aware overlapping constraints that govern how commodity flows relate to actual information flows. 
\end{remark}

\begin{remark}
    Another important novel aspect not yet considered in existing literature is the quantification of the end-to-end service latency violations. We remark that while recent work on VNE~\cite{munk2021good} could be used to provide multi-criteria approximation factors for latency violations at the individual commodity level, the present work is the first to provide approximation factors with respect to {\em end-to-end service service latency} constraints, taking into account both sequential and parallel paths in the service graph (see eqs. \eqref{eq:local_latency}-\eqref{eq:cumulative_latency_total}).
\end{remark}


We start by defining the following set of random variables resulting from the randomized embedding selection process in Step 4 of \alg. 

\begin{definition}
\label{randomvariables1}    
The Random Commodity Flow of $k\in\mathcal K^T$ over link $(u,v)\in\mathcal E^a$ is the Bernoulli random variable: $\flR_{uv}^k \sim \mathcal B(q_k)$
where:
$$q_k = \sum_{n=1}^{N_\phi} p^\phi_n  1\{  \mathsf{f}_{uv}^k(\Emb^{\phi}_n) \neq 0 \} , \text{  with  } \phi  \text{ such that} \,  k \! \in  \! \mathcal K^{T,\phi}.
$$ 
(Recall that a commodity $k\in\mathcal K^{T}$, defined as an edge of in $\mathcal K^T$, can only belong to one connected component $\mathcal R^{T,\phi}$.)
\end{definition}

\begin{definition}
\label{randomvariables2}
The Random Information Object Flow for object $o\in\mathcal O$  over link $(u,v) \in \mathcal{E}^{a}$  is the random variable defined as: 
$$\displaystyle \muR_{uv}^o = {\max_{k \in g_T^{-1}(o)}\left \{R_{uv}^k \flR_{uv}^k  \right\}}$$
\end{definition}

\begin{definition}\label{randomi_inf}
The Random Information Flow $\muR_{uv}$ is defined as the sum over all objects of the random information object flow:
$$
\muR_{uv} = \sum_{o \in \mathcal O} \muR^o_{uv} \qquad \forall (u,v) \in \mathcal E^a
$$
\end{definition}

 \begin{definition}
\label{eq:randomvariables3}
The Random Resource Cost is the random variable defined as 
$$ 
{\CostR}=\sum_{(u,v)\in \mathcal{E}^a} {\muR}_{uv}w_{uv} 
$$
\end{definition}


\begin{definition}
\label{randomvariables5}
The Random Local Latency of commodity $k \in \mathcal{K}^T$ is a random variable defined as: 
$$ \latencyR^{k} = \sum_{(u,v) \in \mathcal{E}^{a}} l^k_{uv} \flR_{uv}^k \qquad\qquad\qquad \qquad  \forall k \in \mathcal{K}^T\label{local_latencyR}$$
\end{definition}


\begin{definition}
\label{randomvariables6}
The Random Cumulative Latency of commodity $k \in \mathcal{K}$ is a random variable defined recursively as: 
$$  \lambda_T^k \geq \lambda^k +\lambda_T^\ell \qquad \qquad\quad \forall k \in \mathcal{K}^{T}
\backslash \mathcal{K}^{s,T}, \ell \in \mathcal{X}^{T}(k)
\label{local_cumulativeR}$$
with  $\lambda^k$ defined in Definition \eqref{randomvariables5}.
\end{definition}





\begin{remark}
Given the disjoint nature of  Algorithm~\ref{disjoint}, i.e. given the fact that Algorithm~\ref{disjoint} decomposes the LP solution for each service tree $\mathcal{R}^{T,\phi} $  independently, and Definition 2, 
 it follows  that 
 the random commodity flows are independent across $k$. 
\end{remark}


We now establish the  connection between the parameter $q_k$ and the  solution to the LP relaxation of MILP \eqref{eq:ilp},  $\{\fLP^k_{u,v}\}$, via Proposition~\ref{lemmaexpCom}.

To this end, we shall first introduce the following Lemma.

\begin{lemma}
\label{decomp}
For any service tree $\mathcal{R}^{T,\phi}$ in $\mathcal R^T$, with $\phi=1, \dots, M$, Algorithm~\ref{disjoint} decomposes the LP commodity flow solution, $\{\fLP^{k}_{uv}\}_{k \in \mathcal K^{T,\phi}}$, into a convex combination of valid mappings, i.e., for all $(u,v)\in\mathcal E^a, k \in \mathcal{K}^{T,\phi}$,
\begin{eqnarray}
\fLP^{k}_{uv}= 
\sum_{n=1}^{N_\phi} 
\flD^k_{uv}(\Emb_n^{\phi}) 
p^\phi_n, \qquad \sum_{n =1}^{N_\phi}p^\phi_n = 1
\end{eqnarray}
with $p^\phi_n$ given in line 17 of  Algorithm~\ref{decomp}.

\end{lemma}

\begin{proof}
The proof leverages the fact that in addition to the fractional commodity flow variables $\{\fLP^k_{uv}\}$, the fractional residual commodity flows computed in Line 18 of Algorithm~\ref{disjoint} and the binary embedding commodity flows, $\flD^k_{uv}(\Emb^{\phi}_n)$, computed in Step 4 of Algorithm~\ref{disjoint} during the $n$-th rounding try, also satisfy the generalized flow conservation constraints \eqref{eq:flowcons}. The complete proof is given in Appendix~\ref{app_lemmadecomp}.
\end{proof}

\begin{proposition}
\label{lemmaexpCom}
For each $ k \! \in  \! \mathcal K^{T}$, the Random Commodity Flow, $\flR^k_{uv}$, defined in Definition \ref{randomvariables1}, is a Bernoulli random variable: $\flR_{uv}^k \sim \mathcal B(q_k)$ with the parameter $q_k$ given by the commodity flow solution of the LP relaxation of MILP \eqref{eq:ilp}, i.e.,
$$
q_k = \fLP_{uv}^k
$$
\end{proposition}
\begin{proof}
From Lemma \ref{decomp} and 
the fact that $\flD_{uv}^k(\Emb^{\phi}_n) \in \{ 1, 0 \}$, it readily follows that 
$\E[\flR_{uv}^k] = q_k= \fLP^k_{uv}$. 
\end{proof}

The next two propositions characterize the random information flow, defined in Definition~\ref{randomvariables2}, when: i) $R_{uv}^k=R_{uv}^o \,\, \forall k \in \{g_T^{-1}(o)\}$, and ii) the rates across commodities representing the same object are not necessarily equal.

\begin{proposition}    
\label{lemmaexpinf}
If $R_{uv}^k=R_{uv}^o$ $\forall k \in \{g_T^{-1}(o) \}$, then the Random Information Object Flow for object $o$ over link $(u,v) \in \mathcal{E}^{a}$, as defined in Definition \ref{randomvariables2}, is a binary random variable taking values in the set $\{0, R_{uv}^o\}$ with probability: 
$$\mathbb{P}(\muR_{uv}^o = R_{uv}^o)= 1-\prod_{k\in  \{g_T^{-1}(o) \}} \left(1-\fLP_{uv}^k\right)$$
\end{proposition}

\begin{proof}
The proof is given in Appendix~\ref{app_lemmaexpinf}.
\end{proof}

\begin{corollary}
Under the conditions of Proposition \ref{lemmaexpinf}, we have:
\begin{align*}
    \mathbb{E}[\muR_{uv}^o]&=
    R_{uv}^{o}\bigg(1-\prod_{k\in g_T^{-1}(o)} (1-\fLP_{uv}^k)\bigg)
\end{align*}
\end{corollary}



\begin{proposition}

\label{lemmaexpinf-2}

The Random Information Object Flow for object $o$ over link $(u,v) \in \mathcal{E}^{a}$, as defined in Definition \ref{randomvariables2}, is a discrete random variable taking values in the set $\bigl\{\{0\} \cup\{R_{uv}^k\}_{k \in g_T^{-1}(o)}\bigl\}$ with probability: 
\begin{eqnarray}
\label{eq:lemmaexpinf2}
    \mathbb{P}(\muR_{uv}^o = R_{uv}^{k_{(i)}})  &=&\prod_{j = i+1}^{|g_T^{-1}(o)|} \left(1- \fLP_{uv}^{k_{(j)}} \right) \fLP_{uv}^{k_{(i)}}
\label{eq:lemmaexpinf3}
\end{eqnarray}
    where 
  $k_{(i)}$ is the i-th commodity, among all commodities representing the same object, ordered in ascending order of their correspondent non-zero rates,   
  i.e., $k_{(1)}$ is the commodity with lowest non-zero rate across all the commodities representing the same object.\footnote{Obviously, 
$\displaystyle \mathbb{P}(\muR_{uv}^o = 0)  =\prod_{j = 1}^{|g_T^{-1}(o)|} \left(1- \fLP_{uv}^{k_{(j)}} \right)$. 
}  
\end{proposition}

\begin{proof}
Proposition \ref{lemmaexpinf-2} follows immediately from Definition \ref{randomvariables1}, Definition \ref{randomvariables2}, and Proposition \ref{lemmaexpCom} after simple algebraic manipulations and recalling that the random commodity flows $\{\flR^k_{uv}\}$ are independent across $k$. 
\end{proof}

\begin{corollary}
Under the conditions of Proposition \ref{lemmaexpinf-2}, we have:
\begin{eqnarray}
\label{eq:lemmaexpinf2}
    \mathbb{E}[\muR_{uv}^o ] \! \!\! \! &=&  \!\!\! \!
  \!\!\!\!\!\!\!  \sum_{i=1}^{|g_T^{-1}(o)|}
    R_{uv}^{k_{(i)}}
\prod_{j = i+1}^{|g_T^{-1}(o)|} \left(1- \fLP_{uv}^{k_{(j)}} \right) \fLP_{uv}^{k_{(i)}}
\label{eq:lemmaexpinf4}
\end{eqnarray}
\end{corollary}

Next, using standard exponential measure concentration bounds, the following theorems provide bounds on the probability of three events: the probability that  the solution of the \alg\, algorithm: 
\begin{enumerate}
\item 
Achieves a total cost that exceeds the optimal objective function value of MILP~\eqref{eq:ilp} (applied to $\mathcal R^T$), $\Cmilp$, by an approximation factor $\Delta_\alpha$.
\item 
Violates the capacity constraint of a given link by a relaxation factor $\Delta_{\beta_1}$.
\item  Violates the cumulative latency constraint of a given destination commodity by a relaxation factor $\Delta_{\beta_2}$.
\end{enumerate}

\begin{theorem}\label{thm1.1}
Let $\beta_1= \Delta_{\beta_1} -  \delta_{\beta_1},
$
where 
$\delta_{\beta_1}= \frac{\mathbb{E}[\muR_{uv}]}{c_{uv}},$ while  $\Delta_{\beta_1}$ is a positive constant larger or equal than 1, such that  $\beta_1>0$ . 
Then: 
$$\mathbb{P}(  \muR_{uv}\geq \Delta_{\beta_1} c_{uv})  \leq  \exp\Bigg[-\frac{2(\beta_1 c_{uv})^2 }{\xi_{uv}}\Bigg]
$$
where
$$\xi_{uv}=\sum_{o \in {\mathcal O}}\left (R^{o,act}_{uv}\right)^2$$
with $R^{o,act}_{uv}= \max_{k\in g_T^{-1}(o)} \left \{ R_{uv}^k 1\{\muLP_{uv}^k>0\} \right \}$. 
\end{theorem}

\begin{proof}
The proof of Theorem \ref{thm1.1} is given in Appendix~\ref{app_thm1}.
\end{proof}

\begin{theorem}\label{thm1.2}
Let  $\beta_2= \Delta_{\beta_2} -  \delta_{\beta_2},
$ with
$\delta_{\beta_2}= \frac{\mathbb{E}[\latencyR^k_{T}]}{L^k}$, 
 and  $\Delta_{\beta_2}$ being a positive constant larger or equal than 1, such that $\beta_2>0$.  Then: 

$$
\mathbb{P}(\latencyR^k_T\geq \Delta_{\beta_2} L^k) \leq \exp\Bigg[-\frac{2(\beta_2 L^k)^2 }{\left(\Lambda^k_{\sf max} - \Lambda^k_{\sf min} \right)^2 }\Bigg]
$$
where $\Lambda^k_{\sf max}$ and $\Lambda^k_{\sf min}$  are defined as follows:
$$\Lambda^k_{\sf max}  \geq \max_{(u,v) \in \mathcal{E}^{a}}  l^k_{uv} {\sf P_k^{\uparrow}}
 +\Lambda_{\sf max} ^\ell \qquad \forall k \in \mathcal{K}^{T} \backslash \mathcal{K}^{s,T}, \ell \in \mathcal{X}^{T}(k)
 $$
 $$\Lambda^k_{\sf min}  \leq  \min_{(u,v) \in \mathcal{E}^{a}} l^k_{uv} {\sf P}_k^{\downarrow}
 +\Lambda_{\sf min} ^\ell \qquad \forall k \in \mathcal{K}^{T} \backslash \mathcal{K}^{s,T}, \ell \in \mathcal{X}^{T}(k)
 $$
 with
 $${\sf P_k^{\downarrow}}= \min_{n} \{ |\map_n^{\mathcal K,\phi}(k) |\}, \qquad {\sf P_k^{\uparrow}} = \max_{n}\{  |\map_n^{\mathcal K,\phi}(k)|\}.$$ 
\end{theorem} 

\begin{proof}
The proof of Theorem \ref{thm1.2} is given in Appendix~\ref{app_thm1}.
\end{proof}

\begin{remark}
Differently from the Random Information Object Flow and
the Random Information Flow,  
the Random Cumulative Latency of a given destination commodity does not admit a simple closed-form expression for its expected value. However, such mean can be efficiently computed using dynamic programming.
\end{remark}

\begin{theorem}\label{thm1.3}
Let $\alpha = \Delta_\alpha- \delta_\alpha,
$ where 
$\delta_\alpha = \frac{\mathbb{E}[\CostR]}{\CostLP}$, while 
$\Delta_\alpha$ is a positive constant larger or equal than 1, such that $\alpha>0$. Then: 
$$
\mathbb{P}\left(\CostR\geq \Delta_{\alpha} \Cmilp\right) \leq \exp\Bigg[-\frac{2(\alpha \CostLP)^2 }{\chi}\Bigg]
$$
where $\Cmilp$ and $\CostLP$ denote the objective function value of MILP~\eqref{eq:ilp} and of its LP relaxation, respectively, 
while 
$$\chi= \displaystyle\sum\limits_{(u,v)\in \mathcal{E}^a}  w_{uv}^2 \left(\sum_{o} R^{o,act}_{uv} \right)^2. $$
\end{theorem} 

\begin{proof}
The proof of Theorem \ref{thm1.3} is given in Appendix~\ref{app_thm1}.
\end{proof}

The following lemma, whose proof is immediate, allows us to provide a tighter multi-criteria approximation: 

\begin{lemma}
\label{lemmascemo}
The probability that 
$\muR_{uv}$ violates the capacity of link $(u,v)
\in \mathcal{E}^{a}$  by a factor $\Delta_{\beta_1}$ is equal to zero 
if condition {\sf F} is satisfied: 
\begin{itemize}
\item \text{{ Condition \sf F}}:  $ \left \{ \, \displaystyle \sum_{o} R^{o,act}_{uv}  \leq \Delta_{\beta_1} c_{uv} \right \}$
\end{itemize}
\end{lemma}
\begin{proof}
The proof follows immediately from the observation that $\sum_{o} R^{o,act}_{uv}$ 
is an upper bound for 
$\muR_{uv}$.
\end{proof}


Based on Lemma \ref{lemmascemo} and  letting  $\mathcal{E}^{\sf F}$  denote 
the set of links for which condition {\sf F} is not satisfied, 
from Theorem \ref{thm1.1}, it follows immediately that
$\mathbb{P}(  \muR_{uv}\geq \Delta_{\beta_1} c_{uv})=0 \quad \forall (u,v) \notin \mathcal{E}^{\sf F}$  
while
\begin{eqnarray}
\mathbb{P}(  \muR_{uv}\geq \Delta_{\beta_1} c_{uv})  &\leq & 
\exp\Bigg[-\frac{2(\beta_1 c_{min})^2 }{\xi_{\sf max}}\Bigg]  
\label{eq:rottaenorme} 
\end{eqnarray}
$\forall (u,v)\in \mathcal{E}^{\sf F}$ with 
$\xi_{\sf max}= \displaystyle \max_{(u,v) \in  \mathcal{E}^{\sf F} }\{\xi_{uv}\}$, and $c_{min}=  \displaystyle\min_{(u,v) \in \mathcal{E}^{\sf F}}c_{uv}$.  

Furthermore, from Theorem~\ref{thm1.2} and letting 
 $$
  \Lambda_{\sf max} \geq \Lambda^k_{\sf max}
  \text{   and   } 
 \Lambda_{\sf min} \leq \Lambda^k_{\sf min},
 \text{  for all  } k \in \mathcal{K}^d, 
 $$
we have that 
\begin{eqnarray}
\mathbb{P}(\latencyR^k_T\geq \Delta_{\beta_2} L^k) \leq  \exp\Bigg[-\frac{2(\beta_2  L_{min })^2}{(\Lambda_{\sf max}-\Lambda_{\sf min})^2 }\Bigg]
\label{saturday1}
\end{eqnarray}
with $L_{min}= \min_{k \in \mathcal K^d}\{L^{k}\}$. 

Using \eqref{eq:rottaenorme} and \eqref{saturday1}, we are now ready to 
quantify the approximation factor, the  relaxation factors, and the probability threshold for the multi-criteria approximation of 
 the \alg\, algorithm.

\begin{definition}
\label{multi-obj2}
Consider an optimization problem with $q\geq 1$ objective functions $h_1(x), h_2(x), \ldots, h_q(x)$ to be minimized  over a feasible set $S$ defined by constraints $g_1(x) - c_1 \leq 0, g_2(x) -c_2 \leq 0, \ldots, g_r(x) -c_r \leq 0$. A solution $x^* \in S$ is a $(\Delta_{\alpha_1}, \Delta_{\alpha_2}, \ldots, \Delta_{\alpha_q}; \Delta_{\beta_1} \Delta_{\beta_2}, \ldots, \Delta_{\beta_r})$-approximation with high probability if there exist approximation factors $\Delta_{\alpha_i}\geq 1$,
relaxation factors $\Delta_{\beta_i}\geq 1$, and  a probability threshold $\epsilon$ close to zero such that:
       \[
        \mathbb{P} \left ( \displaystyle h_i(x^*) \leq \Delta_{\alpha_i} \cdot \min_{x \in S} h_i(x)  \right) \geq 1- \epsilon \quad  \forall  i = 1, 2, \ldots, q  \]

and 
\[
      \mathbb{P} \left(g_j(x^*) \leq \Delta{\beta_j} \cdot c_j \right) \geq 1- \epsilon  \quad \forall j = 1, 2, \ldots, r
\]
\end{definition}









\begin{theorem} 
\label{thm2}
After $t$ rounding tries, with probability $1-\theta^t$ with $\theta<1$, 
the \alg \, algorithm returns, with high probability,  a $(\Delta_\alpha;\Delta_{\beta_1}, \Delta_{\beta_2})$-approximation 
for the \prob\, problem  in \eqref{eq:ilp},
i.e. 
\begin{eqnarray} 
\mathbb{P}(\CostR \leq \Delta_{\alpha} \Cmilp) & \leq &  1- \theta^t,    \\
\mathbb{P}(  \muR_{uv}\leq \Delta_{\beta_1} c_{uv})  & \leq &  1- \theta^t,      \quad \forall (u,v) \in \mathcal E^a,  \\
\mathbb{P}(\latencyR^k_T \leq   \Delta_{\beta_2} L^k)   & \leq &  1- \theta^t,
\quad  \forall k \in \mathcal K^d, 
\end{eqnarray}
where:
$$
\Delta_\alpha = \frac{1}{\CostLP}\sqrt{\chi\frac{\ln(\theta/3)}{2}} + \delta_\alpha
$$

$$
\Delta_{\beta_1} = \kappa \sqrt{\frac{1}{2}\ln \left (\frac{\theta |\mathcal{E}^{\sf F}|}{3} \right)} 
     + \max_{(u,v)\in \mathcal{E}^{\sf F}} \left\{  \frac{\mathbb{E}[\muR_{uv}]}{c_{uv}}
\right\}
$$
and
$$
\Delta_{\beta_2} = \sqrt{\frac{1}{2}\ln \left (\frac{\theta |\mathcal{K}^d|}{3} \right) }  \frac{(\Lambda_{\sf max}-\Lambda_{\sf min}) }{(L_{min })} + 
\max_{k\in \mathcal{K}^{d}} \left\{  \frac{\mathbb{E}[\latencyR^k_{T}]}{L^k} \right\}
$$

\end{theorem}

\begin{proof}
The proof of Theorem \ref{thm2} is given in Appendix~\ref{app_thm2}.
\end{proof}

\section{Practical Extensions}\label{sec:extensions}

\subsection{Resource Allocation}
\label{sec:res_allocation}

We remark that, for ease of exposition, we have started with a formulation of the \prob\, in \eqref{eq:ilp} that assumes a flow-proportional cost model determined by the cost per unit flow of operating a given resource. While this model is being increasingly used by e.g., cloud providers in their more elastic 
FaaS (Function as a Service) compute-models, in many cases, cost models that charge for the use of discrete resource blocks are also of practical relevance. 
Examples 
may include communication resource blocks, (e.g., time-frequency blocks, wavelengths), computation resource blocks (e.g., CPUs), and storage/memory resource blocks (e.g., RAM modules, disks).  
In addition, computation blocks with predefined CPU and memory configurations can also be used, e.g., containers or virtual machines (VMs).

Incorporating resource blocks into our formulation can be achieved by introducing the following variables:
\begin{itemize}
    \item $c_{uv}^T$: the total number of blocks that can be allocated to link  $(u,v)$.
    \item $c^b_{uv}$: the capacity per block (in flow units, e.g., Mbps) at link $(u,v)$.
    \item $w^b_{uv}$: the cost per allocated block at link $(u,v)$.
    \item $y_{uv}$: the number of resource blocks allocated to link $(u,v)$.
\end{itemize}

The objective function is then modified as follows:
\begin{eqnarray}
\text{min}\,\,\displaystyle\sum\limits_{(u,v)\in \mathcal{E}^a} y_{uv}w^b_{uv}
\label{eq:Resource_Allocation}   
\end{eqnarray}

Furthermore, the following two constraints that guarantee that the total rate at link $(u,v)$ is covered by enough resource blocks, and that the number of allocated resource blocks does not exceed the maximum number of available blocks, are added:
\begin{align}
\mu_{uv} \leq y_{uv}c_{uv}^b \qquad \forall (u,v) \in \mathcal{E}^a  \label{eq:cap_resource} \\
y_{uv} \leq c_{uv}^T \qquad \forall (u,v) \in \mathcal{E}^a \label{eq:cap_blocks} 
\end{align}



Note that these modifications to the original formulation do not change the solution provided by the associated LP relaxation. 
In fact, denoting by $\yLP_{uv}$
the resource allocation solution of the LP relaxation, 
we have 
\begin{eqnarray}
\label{eq:resource_block_LP}
\yLP_{uv}=\frac{\muLP_{uv}}{c_{uv}^b}. 
\end{eqnarray} 

Consequently, the version of  the \alg\, 
 \,algorithm that incorporates
 resource blocks follows the same steps described in Sec. \ref{sebsectionIDAGO}, 
 by just adding at the end of  Step 5, the  computation of  the number of resource blocks for link $(u,v)$ as follows: 
$$
\yD_{uv}(\underline{\Emb}) = \left \lceil \frac{\muD_{uv}(\underline{\Emb})}{c_{uv}^b} \right  \rceil.
$$

Therefore, the performance analysis of \alg\, with resource blocks 
follows a very similar analysis to that of Sec. \ref {sec:analysis}.
Specifically, let us first introduce the following definition: 
\begin{definition}
\label{randomRes_block} 
The Random Resource  Blocks $\yR_{uv}$ is defined as:
$$
\yR_{uv}
= \left \lceil \frac{\muR_{uv}}{c_{uv}^b} \right  \rceil. 
$$
\end{definition}

Starting from Definition \ref{randomRes_block} and using Eq.  \eqref{eq:resource_block_LP}, 
it is possible to bound the probability that the number of resource blocks allocated to link $(u,v)$ exceed the maximum number of available blocks, leading to the following corollary. 


\begin{corollary}
\label{corrollary_resous_block}
    The probability that the random resource blocks, $\yR_{uv}$, allocated, by the \alg\, algorithm,
    to edge $(u,v)\in \mathcal{E}^{a}$ exceeds the maximum number of available blocks $c^T_{uv}$ by a factor $\Delta_{\beta_1}$ can be bounded as follows:
$$
\mathbb{P}(\yR_{uv} \geq \Delta_{\beta_1} c^T_{uv})  \leq \exp\Bigg[-\frac{2(\beta_1 c_{uv}^T -1)^2 }{\xi_{uv}^T}\Bigg] 
$$
with  $\Delta_{\beta_1}$  and  $\beta_1$ defined as in Theorem \ref{thm1.1}.

\end{corollary}
\begin{proof}
The proof of Corollary \ref{corrollary_resous_block} is given in Appendix~\ref{app_corrollary_resous_block}.
\end{proof}
\subsection{Service Dynamics}
In line with multi-scale orchestration solutions such as~\cite{pagliuca2024dual}, we envision end-to-end service optimization algorithms like \alg\, running at centralized controllers that operate at a longer timescale 
and can hence leverage global network view to optimize end-to-end service distribution, while complemented with distributed control policies operating at a shorter timescale 
to adjust resource allocation decisions based on local real-time observations.

In this context, the goal of the resource allocation solution provided by the long-term optimization algorithm is to {\em cover} the total {\em average} information flow rate, leaving the short-term control policies to handle the {\em instantaneous} rate variations (due to the stochastic nature of service rates) via dynamic resource autoscaling mechanisms (e.g., Kubernetes microservice autoscaling~\cite{kubernetes2020kubernetes}). 

Nonetheless, in practice, it may still be relevant for the long-term optimization algorithm to provide a resource allocation solution that covers the average flow rate with a certain {\em margin}, whose value should be driven by the dynamics of the service rates and the availability of resource autoscalers. In fact, such an approach tries to strike a balance between traditional over-provisioning approaches that make sure peak rates are covered at the expense of excessive resource waste, and autoscaling approaches that try to follow the instantaneous rate at the expense of extra orchestration cost.

To this end, we introduce the \textit{burstiness factor} of commodity $k$ over link $(u,v)$, denoted by $B^k_{uv}$, in order to capture (i) the dynamics of commodity $k$ as well as (ii) the availability of dynamic resource autoscalers at link $(u,v)$. 
We can then  extend the \prob\, problem formulation in \eqref{eq:ilp} to incorporate the burstiness factor $\burfact$ by simply adjusting Eq. \eqref{eq:mult1} as follows:
\begin{align}
f_{uv}^k R_{uv}^k \burfact \leq \mu_{uv}^o  \qquad \forall (u,v) \in \mathcal{E}^a, k \in \mathcal{K}, o = g(k).\notag
\end{align}



\begin{figure}[h]
    \centering
    \includegraphics[width=0.9\linewidth]{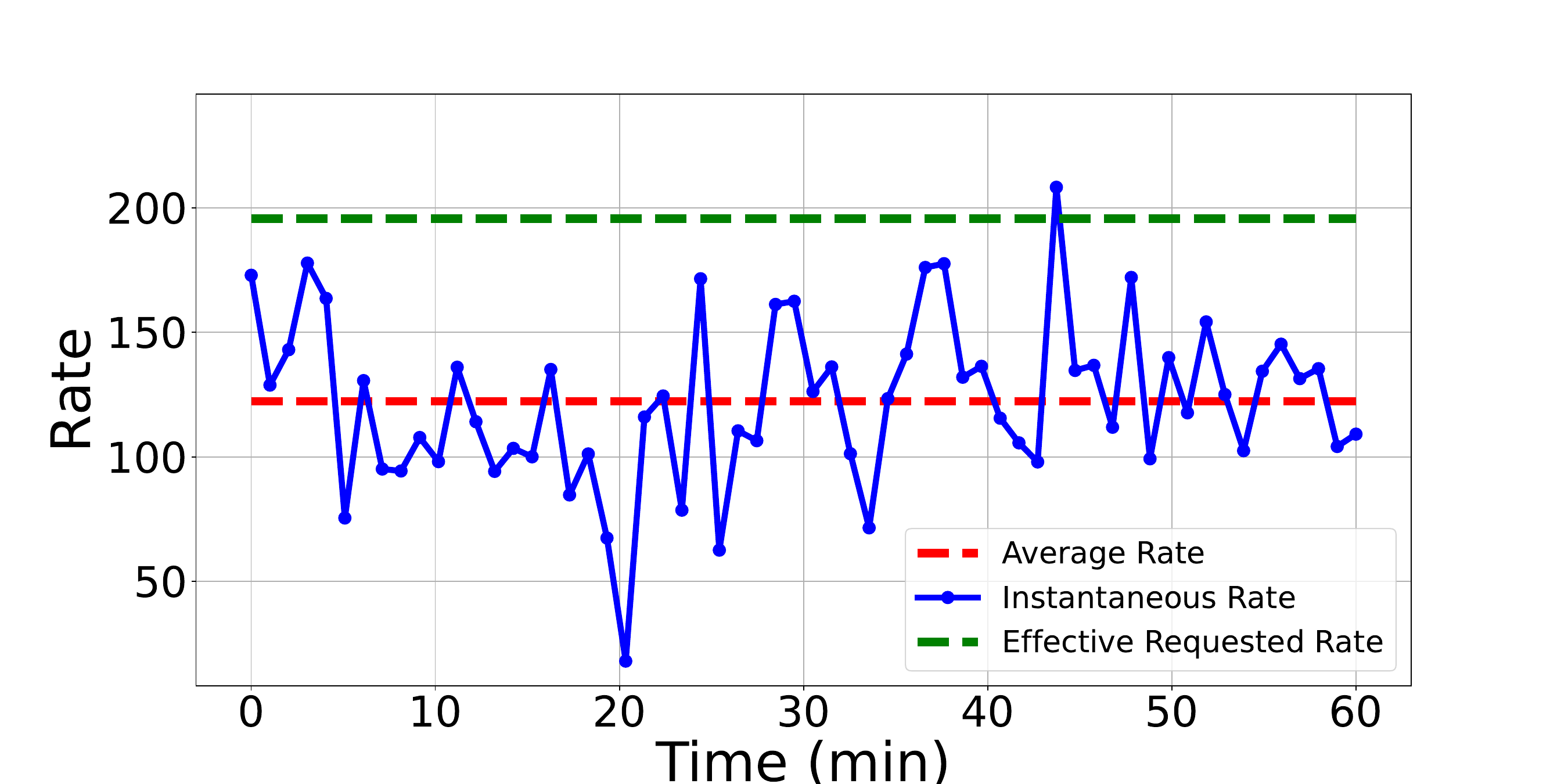}
    \caption{\small{Illustration of instantaneous rate, average rate, and effective requested rate.}}
    \label{fig:burstiness}
\end{figure}

Fig.~\ref{fig:burstiness} illustrates the concept of the burstiness factor for long-term resource allocation. 
The blue line shows the instantaneous rate of a given commodity $k$ on link $(u,v)$ over a given service session. 
The red dashed line represents the average rate, $R^k_{uv}$, while 
the green dashed line depicts the {\em effective requested rate}, i.e., the product of the average rate times the burstiness factor $\burfact$.


\section{Evaluation Results}
\label{results}
In this section, we evaluate the performance of the proposed \alg\, algorithm through extensive simulations in a variety of network settings and NextG media service configurations. 

In {\em Scenario 1}, we focus on smaller-scale network and service settings using easy-to-grasp synthetic data in order to clearly illustrate the benefits of \alg's main innovations: (i) the capability of the 
DAG-to-Forest service graph transformation to maximize replication opportunities, and (ii) the efficacy of the resulting relaxation, decomposition, and rounding process to provide tight approximation guarantees.

In {\em Scenario 2}, we then focus on illustrating the practical cost reductions obtained in the context of a larger-scale network with realistic resource cost and VR application data.

In terms of benchmarking, we consider the following state-of-the-art solutions:
\begin{itemize}
    \item {\em MILP Info-Unaware DAG}: exponential-time solution obtained solving MILP~\eqref{eq:ilp} using original service DAG, but treating each commodity as a different information object, i.e., akin to VNE. 
    \item {\em MILP Info-Aware DAG}: exponential-time solution obtained solving MILP~\eqref{eq:ilp} using original information-aware service DAG. 
    \item {\em MILP Info-Aware Forest}: exponential-time solution obtained solving MILP~\eqref{eq:ilp} using  information-aware transformed service Forest. 
    \item {\em \alg}: polynomial-time solution obtained by the proposed \alg\, algorithm.
\end{itemize} 


For \alg's approximation performance evaluation,  
we concentrate on the following metrics closely related to Theorems~\ref{thm1.1} and~\ref{thm1.2}.\footnote{
Note that Theorems \ref{thm1.1} and \ref{thm1.2} can be formulated in terms of the random cost approximation ratio, defined as $\frac{\CostR}{\Cmilp}$, and  the random capacity relaxation factor, defined as $ \max_{(u,v) \in \mathcal{E}^a} \left \{ \frac{\muR_{uv}}{c_{uv}} \right \}$.}
 
\begin{itemize}
    \item \textbf{Cost Approximation Ratio (\textit{\ratio})}:
    The ratio between the cost obtained by \alg\, and the cost of MILP \eqref{eq:ilp} on the information-aware Forest: 
    \[
    \ratio = \frac{\CostIDAGO}{\Cmilp}
    \]   
    \item \textbf{Capacity Relaxation Factor (\textit{\nwuse})}: 
    The maximum over all links, of the ratio between the link information flow obtained by \alg\, and its capacity:
    \[
    \nwuse = \max_{(u,v) \in \mathcal{E}^a} \big(\frac{\muD_{uv}(\underline{\Emb})}{c_{uv}}\big)
    \]   
\end{itemize}

\subsection{Scenario 1 - Setting A: Low congestion}

In Scenario 1, Setting A, we evaluate the 
benefit of \alg's DAG-to-Forest service graph transformation procedure to maximize replication opportunities. 

\begin{figure}[h]
    \centering
    \includegraphics[width=0.85\linewidth]{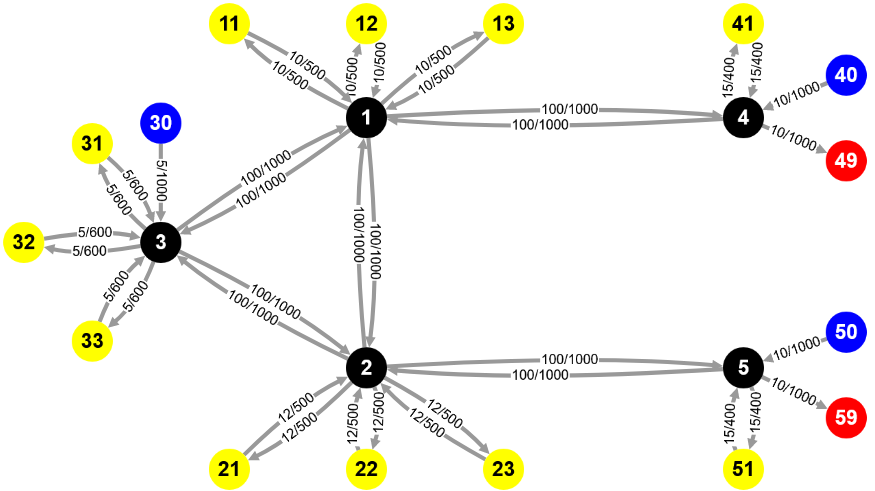}
    \caption{\small{Cloud-augmented graph in Scenario 1.}}
    \label{fig:network_scenarioA}
\end{figure}

\subsubsection{\textbf{Network configuration}}\label{sec:network_config_scenarioA}
We consider a hierarchical network composed of one core cloud node, two edge cloud nodes, and two compute-enabled access points, as depicted in Fig.~\ref{fig:network_scenarioA}. In the cloud-augmented network, black circles represent communication nodes, yellow circles represent computation clusters available at the corresponding core/edge/access nodes, blue and red circles represent source and destination endpoints, respectively. The numbers on the cloud-augmented graph links indicate associated resource costs and capacities (cost/capacity). Table \ref{tab:network_config_scenA} provides detailed information on the capacity and cost of each network link.
We recall that the appropriate capacity units vary depending on the type of resource represented by each network link. The capacity of communication links is measured in communication flow units (e.g., Mbps), the capacity of "computation out" links in computation flow units (e.g., Gflops), and the capacity of "computation in" links in "storage flow" units (e.g., MB). The resource costs are given in cost per unit flow per hour, indicating the operational expenses (OPEX) associated with running a given resource during an hour per flow unit.


\begin{table}[h]
    \centering
        \begin{tabular}{|c|c||c|}
        \hline RESOURCE & CAPACITY & UNIT COST per HR\\
        \hline
        
        \hline Core$\leftrightarrow$Edge Links & $500$ Mbps & $100$ \\
        \hline Edge1$\leftrightarrow$Edge2 Link & $500$ Mbps& $100$ \\
        \hline Edge$\leftrightarrow$Access Links & $500$ Mbps& $100$ \\
        \hline Core Computation & $600$ Gflops / $600$ MB&$5$\\
        \hline Edge1 Computation & $500$ Gflops / $500$ MB&$10$\\
        \hline Edge2 Computation & $500$ Gflops / $500$ MB&$12$\\
        \hline Access Computation & $400$ Gflops / $400$ MB & $15$\\
        
        
        
        
        \hline
        \end{tabular}
    \caption{\small{Cloud-network capacity and cost parameters for Scenario 1.}}
    \label{tab:network_config_scenA}
\end{table}

\subsubsection{\textbf{Service configuration}}\label{sec:service_config_scenarioA}

We consider a generic yet representative service graph model that mirrors structures commonly observed in NextG media applications, typically composed of the following three processing stages: 
\begin{itemize}
    \item \textbf{Tracking} (context understanding): Analyze source sensor outputs to understand users' context, intentions, and actions.
    \item \textbf{Synthesis} (experience composition): Compose user experience accessing relevant content from the \textbf{Content Store} and putting it together according to the Tracking output.
    \item \textbf{Personalization}: add user-specific elements and render overall multimedia experience (e.g., XR rendering). 
\end{itemize}





An example of such a NextG media service graph is depicted in Fig.~\ref{fig:original_fac10_scenA}a, where we assume source sensor data from two user groups must be processed by a joint Tracking function, a Synthesis function (with access to a content store), and two Personalization functions. Service commodity rates (production, communication, consumption)\footnote{Recall that in CNFlow, a given commodity is associated with three types of rate, depending on the link (resource) it goes through, i.e, production rate, communication rate, and consumption rate.} are described in Table \ref{tab:service_config_cmds_scenA}, with communication rates also depicted on the edges of the service graph. 


We note that source data streams will also need to be processed by network functions (NFs) before they reach the application functions, 
and our model can easily add NFs to the orchestration problem by including them in the service graph. 
In this evaluation, since our focus is on the application, we use the next-generation node (gNB) NF as the effective source and destination functions of our service graph, acting as ingress and egress points for the application data streams. 

\begin{table}[h]
    \centering
        \begin{tabular}
        {|p{1.1in}|p{0.5in}|p{0.6in}|p{0.5in}|}
        \hline Commodity & Prod. Rate (Gflops) & Comm. Rate (Mbps) & Cons. Rate (MB) \\
        \hline (gNB1, Tracking) & $10$ & $10$ & $10$ \\
        \hline (gNB2, Tracking) & $10$ & $10$ & $10$ \\
        \hline (CS, Synthesis) & $10$ & $10$ & $10$ \\
        \hline (Tracking, Synthesis) & $5$ & $5$ & $5$ \\
        \hline (Synthesis, Pers1) & $15$ & $15$ & $15$ \\
        \hline (Synthesis, Pers2) & $15$ & $15$ & $15$ \\
        \hline (Pers1, gNB1) & $20$ & $20$ & $20$ \\
        \hline (Pers2, gNB2) & $20$ & $20$ & $20$ \\
        \hline
        \end{tabular}
    \caption{\small{Service commodity rates for Scenario 1.}}
    \label{tab:service_config_cmds_scenA}
\end{table}




\subsubsection{\textbf{Results}}

We evaluate the performance of \alg\, computing the total cost obtained by the different approaches described in Sec. \ref{results} as a function of a rate scaling factor that accounts for rate increases driven by factors such as higher service quality/resolution, increased number of users, and/or higher service dynamics (burstiness).



\begin{figure}
    \centering
    \includegraphics[width=\linewidth]{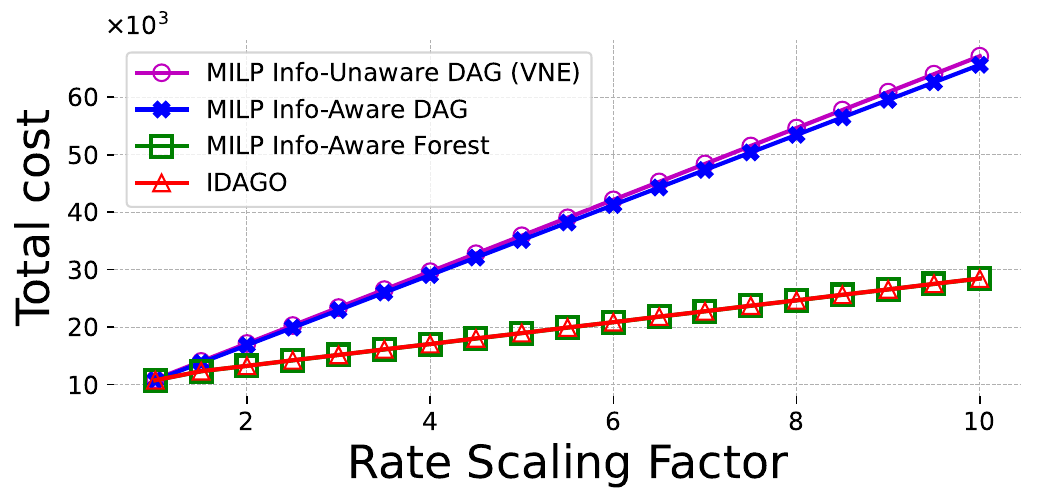}
    \caption{\small{Total resource cost vs rate scaling factor for Scenario 1 - Setting A.}}
    \label{fig:total_cost_scenA}
\end{figure}

In particular, Fig.~\ref{fig:total_cost_scenA} shows the total cost (including computation, communication, and storage resource costs) vs the service rate scaling factor that multiplies the rate of the Synthesis and Personalization outputs. 

As observed in Fig.~\ref{fig:total_cost_scenA}, the VNE-based solution that results from solving \eqref{eq:ilp} on the information-unaware DAG yields the highest cost. While using information awareness on the original DAG allows a slight improvement in cost, the largest cost reduction is obtained when using the information-aware Forest, highlighting the benefit of the DAG-to-Forest transformation procedure to maximize the cost savings brought by efficient flow/function replication. Note also how in this case, \alg\, is able to obtain the optimal solution. In fact, due to the less capacity-constrained conditions of Setting A, \alg\, is able to compute a single embedding that matches the optimal solution to MILP \eqref{eq:ilp} in polynomial time.

\begin{figure}[htbp]
  \centering
  
  \begin{subfigure}[(a)]{0.7\linewidth}
    \centering
    \includegraphics[width=\linewidth]{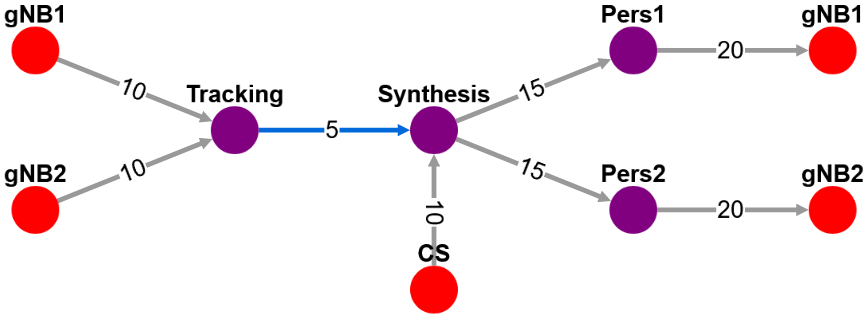}
    \caption{}
    \label{fig:OG_tracking_settingA}
  \end{subfigure}
  
  \begin{subfigure}[(b)]{0.7\linewidth}
    \centering
    \includegraphics[width=\linewidth]{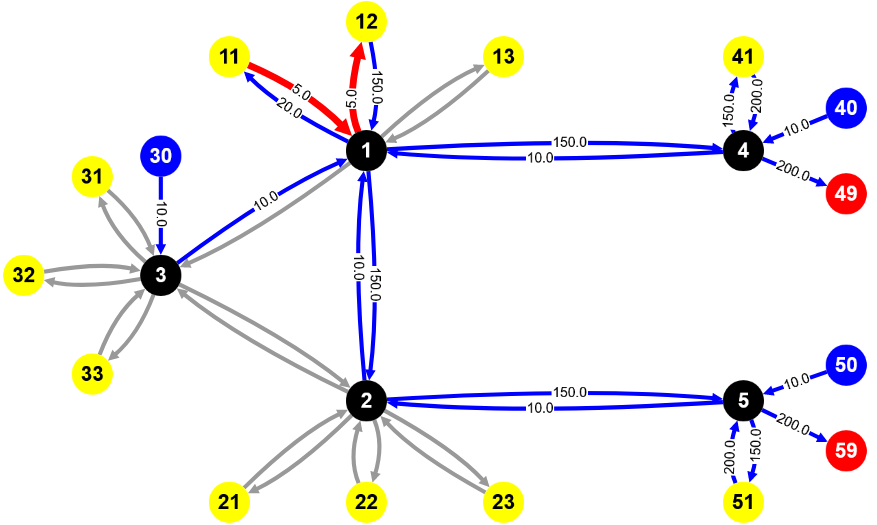}
    \caption{}
  \end{subfigure}
  
  \caption{\small{(a) Original service DAG. (b) Associated flow solution (in red, information flow of Tracking output).}}
  \label{fig:original_fac10_scenA}
\end{figure}

\begin{figure}[htbp]
  \centering
  
  \begin{subfigure}[a]{0.7\linewidth}
    \centering
    \includegraphics[width=\linewidth]{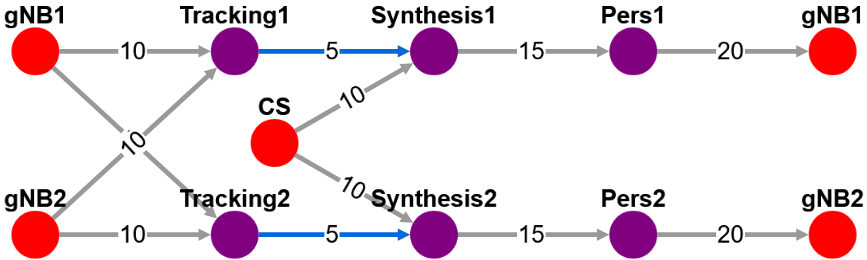}
    \caption{}
  \end{subfigure}
  
  \begin{subfigure}[b]{0.7\linewidth}
    \centering
    \includegraphics[width=\linewidth]{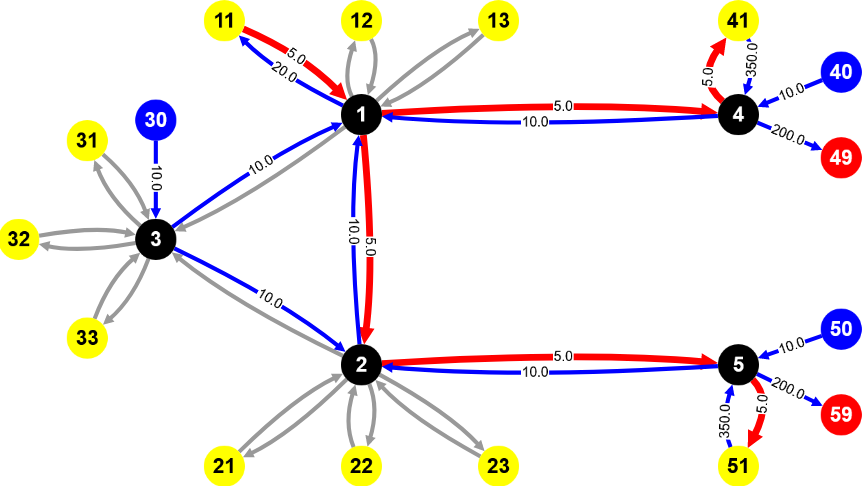}
    \caption{}
  \end{subfigure}
  
  \caption{\small{(a) Transformed service forest. (b) Associated flow solution (in red, information flow of Tracking output).}}
  \label{fig:transf_fac10_scenA}
\end{figure}

To further illustrate the key benefit of the DAG-to-Forest transformation, we depict the flow solution for MILP info-aware DAG and MILP info-aware Forest 
in Figs.~\ref{fig:original_fac10_scenA} and \ref{fig:transf_fac10_scenA}, respectively (for a rate scaling factor of 10).\footnote{In this section, for ease of exposition and noting that the last step of the DAG-to-Forest transformation does not alter the nature of the transformed graph due to source functions being associated with fixed network locations, we depict the transformed graph without the last step.} The information flow solutions are shown in blue on the respective cloud-augmented graphs, with link numbers indicating the resulting total load. For clarity, we also mark in red the flow solution associated with the output of the Tracking function. Observe how in the MILP-DAG solution of Fig.~\ref{fig:original_fac10_scenA}, the output of the Tracking is processed by the Synthesis function at edge node, with Personalization functions 1 and 2 running at the accessed nodes, closer to the end users. However, in the MILP-Forest solution shown in Fig.~\ref{fig:transf_fac10_scenA}, the Tracking output is already replicated and processed at the access nodes, significantly reducing communication cost. 

\subsection{Scenario 1 - Setting B: High congestion}

Setting B expands Setting A by considering a larger service graph composed of multiple connected components, with the goal of assessing the scalability and performance of the proposed framework under increased loads.  




  

\begin{figure}[htbp]
  \centering
  
  \begin{subfigure}[a]{0.67\linewidth}
    \centering
    \includegraphics[width=\linewidth]{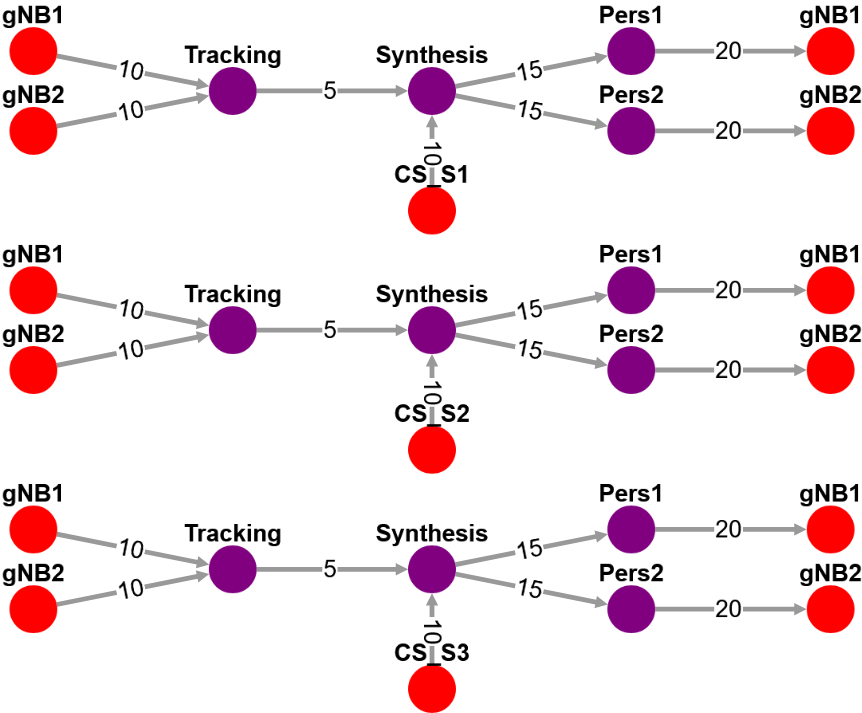}
    \caption{}
    \label{fig:og_settB}
  \end{subfigure}
  
  \begin{subfigure}[b]{0.67\linewidth}
    \centering
    \includegraphics[width=\linewidth]{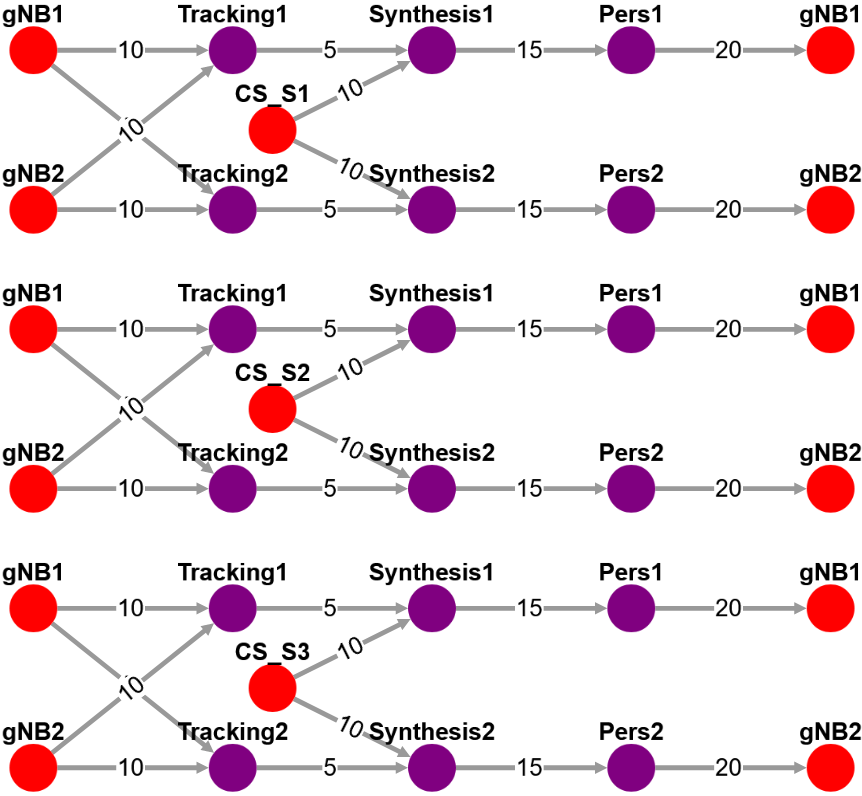}
    \caption{}
    \label{fig:tg_settB}
  \end{subfigure}
  
  \caption{\small{(a) Original service DAG. (b) Transformed service forest.}}
  \label{fig:initial_state_scenarioB}
\end{figure}


\subsubsection{\textbf{Network Configuration}}
\label{sec:network_config_scenarioB}
We use the same cloud-network as in Setting A, described in Fig.~\ref{fig:network_scenarioA} and Table \ref{tab:network_config_scenA}.

\subsubsection{\textbf{Service configuration}}
\label{sec:service_config_scenarioB}
We consider a service collection consisting of three connected components, each representing an instance of a NextG media service of the type described in Setting A, consumed by different user groups, as shown in 
Fig.~\ref{fig:og_settB}. 
This setup increases overall cloud-network resource load, allowing for the evaluation of the \alg\, algorithm in high congestion settings. 
The corresponding DAG-to-Forest transformation is illustrated in Fig.~\ref{fig:tg_settB}. 
The commodity rates are in line with Setting A and described in Table \ref{tab:service_config_cmds_scenA}.

\subsubsection{\textbf{Results}}

\begin{figure*}[h]
    \centering
    \begin{minipage}[b]{0.3\textwidth}
        \centering
        \includegraphics[width=\textwidth]{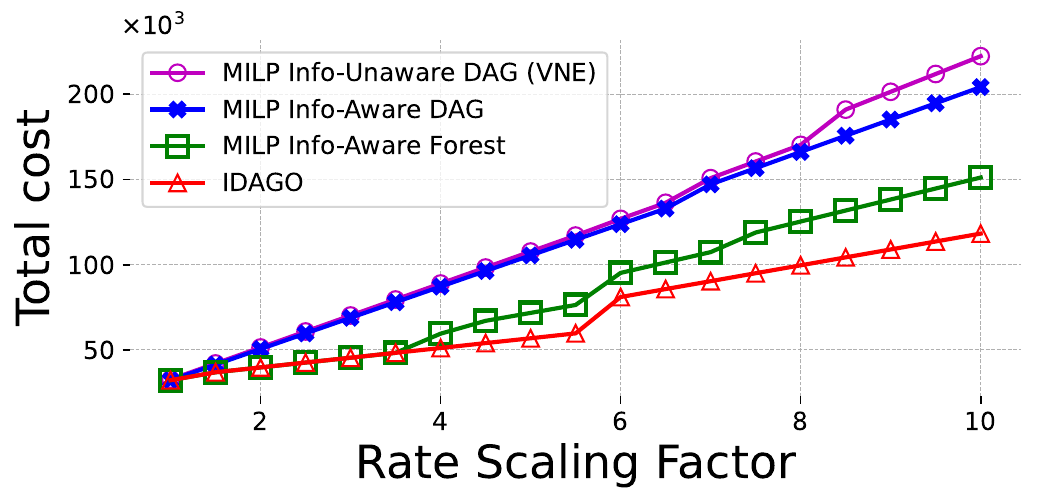}
        \subcaption{}
        \label{fig:total_cost_scenB}
    \end{minipage}
    \hfill
    \begin{minipage}[b]{0.3\textwidth}
        \centering
    \includegraphics[width=\textwidth]{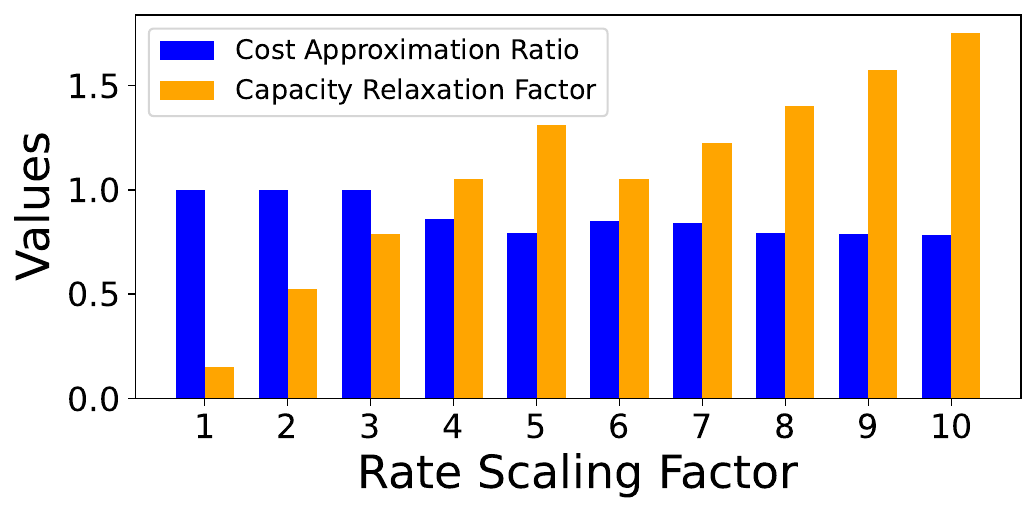}
    \subcaption{}
    \label{fig:alfas_betas_scenB}
    \end{minipage}
    \hfill
    \begin{minipage}[b]{0.3\textwidth}
        \centering
    \includegraphics[width=\textwidth]{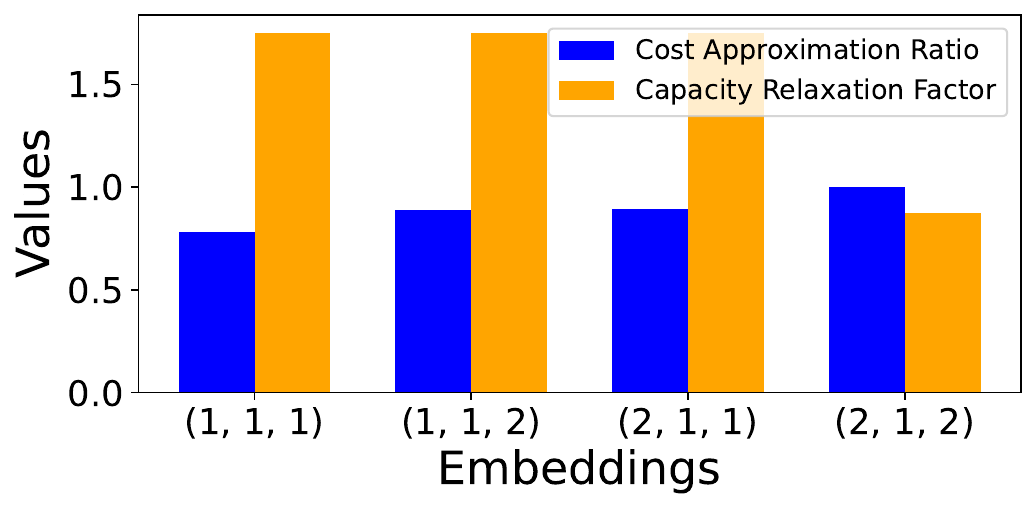}
    \subcaption{}
    \label{fig:alfas_betas_disjoint10_scenB}
    \end{minipage}
    \caption{\small{(a) Total cost vs rate scaling factor. (b) \ratio\, and \nwuse\, for each rate scaling factor. (c) \ratio\, and \nwuse\, values for each embedding $\Emb^{\phi}$ with a rate scaling factor equals to 10.}}
\end{figure*}

Fig.~\ref{fig:total_cost_scenB} shows the total cost vs the service rate scaling factor for \alg\, and baseline solutions. In addition, Fig.~\ref{fig:alfas_betas_scenB} shows the CAR and CRF values associated with the \alg\, solution. 
For \alg, Figs.~\ref{fig:total_cost_scenB} and~\ref{fig:alfas_betas_scenB} show the cost and approximation factors associated with the embedding that yields the lowest cost. 
In line with Setting A, we can observe the progressive cost reductions achieved by taking into account information-awareness on the original DAG, but especially via the DAG-to-Forest graph transformation. 

Focusing on \alg, note how starting from a rate scaling factor of 4, it achieves a total cost even lower than the optimal, at the expense of capacity violations, as illustrated by the CAR and CRF values in Fig.~\ref{fig:alfas_betas_scenB}. 

As previously mentioned, the total cost, \ratio\, and \nwuse\, value in Figs.~\ref{fig:total_cost_scenB} and \ref{fig:alfas_betas_scenB} relate to the embedding that yielded the minimum cost. Nevertheless, in this higher-congestion setting, \alg\, generated 4 embeddings, each with different CAR and CRF values. Fig.~\ref{fig:alfas_betas_disjoint10_scenB} illustrates the \ratio\, and \nwuse\, values of all embeddings generated by \alg\, for a scaling factor 10. Recall that the entire set of embeddings generated for $\miR^T$ results from the Cartesian product of the embeddings of each component $\miR^{T,\phi} \in \miR^T$. In this case, \alg\, obtained 2 embeddings for the first and third components, and 1 for the second. Observe how the choice of embedding can be driven by cost vs capacity violation preferences. In this setting, for example, embedding $(2,1,2)$, with CAR=1 and CRF$<$1, would be the preferred choice to avoid capacity violations, and it is also the optimal solution to MILP \eqref{eq:ilp}. In practice, one can tune the {\em return condition} after Step 5 of \alg\, to decide when to stop the rounding tries according to the customer preferences.


\subsection{Scenario 2: Realistic VR and network data}


With the goal of evaluating \alg\, in closer to realistic settings, we now include (i) the practical aspects described in the extensions outlined in Sec. \ref{sec:extensions}, as well as (ii) realistic network and VR application data:
\begin{itemize}
    \item \textbf{Resource Blocks}: We incorporate discrete resource blocks for computation, communication, and storage resource allocation.
    \item \textbf{Burstiness Factor}: We incorporate the burstiness factor into the rate scaling factor to take into account the stochastic nature of service rates.
    \item \textbf{Practical Cloud-Network}: We integrate infrastructure cost and capacity values from leading cloud providers such as AWS.
    \item \textbf{Practical VR Application}: We utilize real-world VR application data. 
\end{itemize}

For the case of VR applications, the generic processing stages of NextG media services considered in Scenario 1 particularize to:
\begin{enumerate}
    \item \textbf{Tracking}: Analyzes data streams from users' sensors to understand context, environment, and users' interactions. 
    \item \textbf{VR Processing}: Processes tracking data in order to compose the user experience by selecting appropriate media objects from the Content Store.
    \item \textbf{Rendering}: Renders the final video frames consumed by the ned users.
\end{enumerate}


\begin{figure*}[h]
    \centering
    \begin{minipage}[c]{0.3\textwidth}
        \centering
        \includegraphics[width=0.8\textwidth]{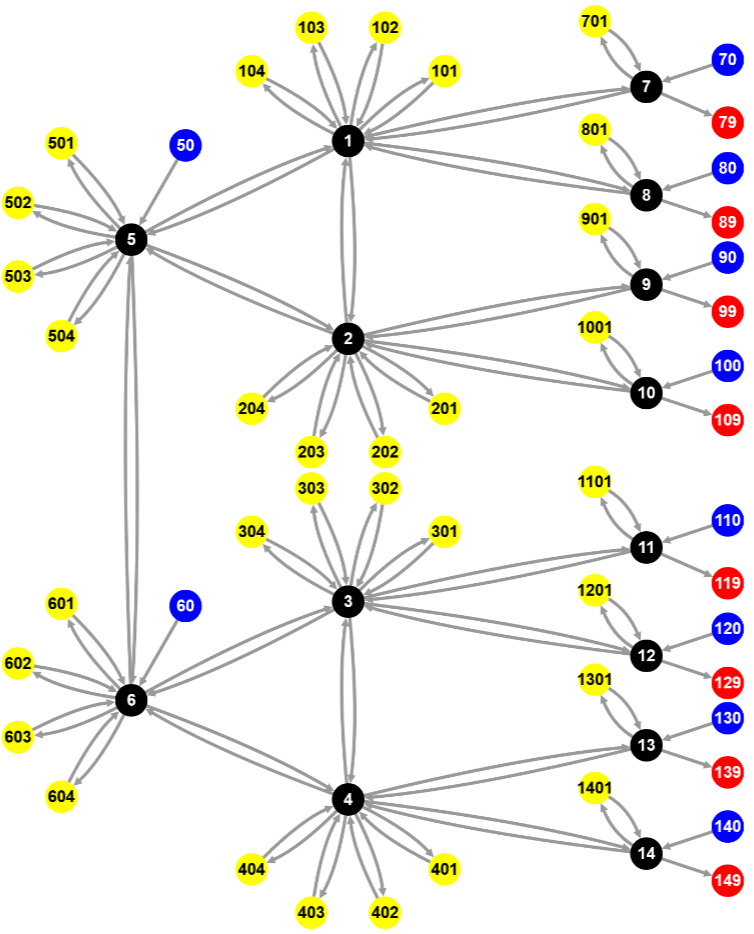}
        \subcaption{}
        \label{fig:nw_scen2}
    \end{minipage}
    \hfill
    \begin{minipage}[c]{0.3\textwidth}
        \centering
    \includegraphics[width=0.8\textwidth]{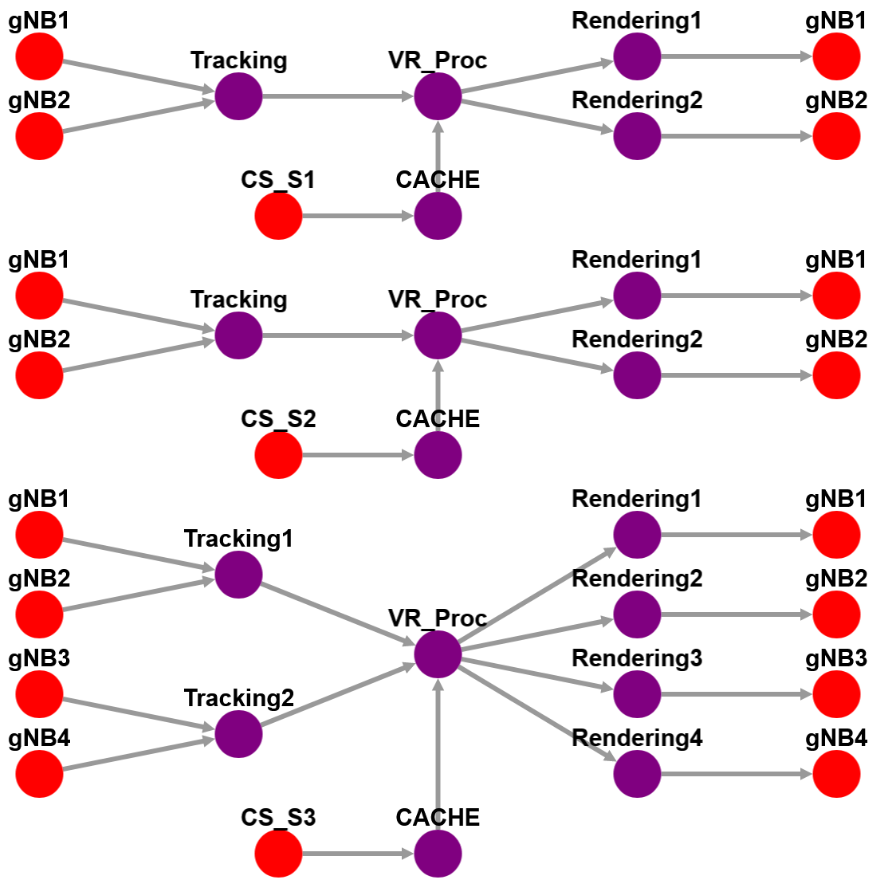}
    \subcaption{}
    \label{fig:og_scen2}
    \end{minipage}
    \hfill
    \begin{minipage}[c]{0.3\textwidth}
        \centering
    \includegraphics[width=0.8\textwidth]{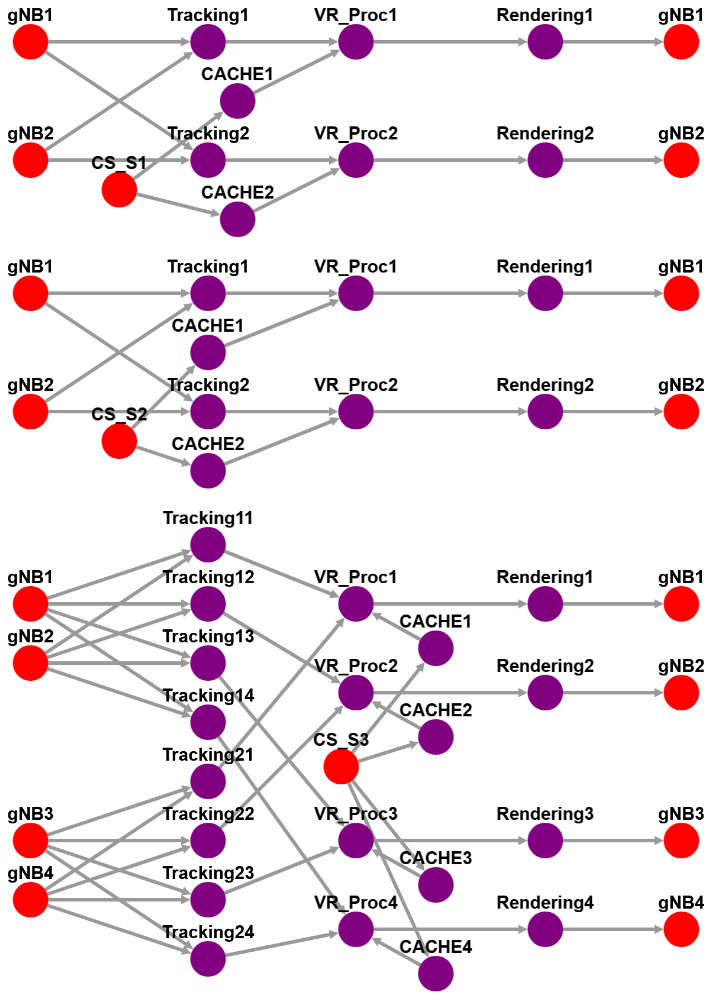}
    \subcaption{}
    \label{fig:tg_scen2}
    \end{minipage}
    \caption{\small{(a) Cloud-network graph. (b) Original service DAG. (c) Transformed service Forest.}}
    \label{fig:initial_state_scenarioC}
\end{figure*}


We assume the Content Store (CS) to be located in the core cloud, but we also consider a Cache function that allows caching frequently used VR objects closer to the user if beneficial.

We remark that while most current VR systems are monolithic and running mostly in cumbersome user devices, 
there is ongoing interest in disaggregating the application to leverage edge/cloud resources e.g., via 
\textit{remote rendering} efforts~\cite{remote-rendering}, where Tracking functionality can be physically separated from 
VR processing and rendering functions. 
Our objective 
is to explore further disaggregation opportunities to assess how flexibly distributing these components over the compute continuum can impact efficiency, scalability, and performance of NextG resource-intensive media applications.\footnote{Note that while disaggregation offers modularity and flexibility, our optimization framework can still decide to colocate service functions if no benefit is identified from distribution.} 

\subsubsection{\textbf{Network configuration}}
The cloud-network topology in Scenario 2 results from duplicating that of Scenario 1, yielding two core cloud nodes, four edge cloud nodes, and two access points per edge node, totaling eight access points. The associated cloud-augmented graph is illustrated in Fig.~\ref{fig:nw_scen2}. 
In this setup, each core node represents an AWS region. Each edge node an AWS edge location, and each access node a compute-enabled access point managed by a local network operator, such as Verizon or AT\&T. We consider AWS Direct Connect prices \cite{AWS_DTO} for communication costs, scaling by $2$X the cost of network operator managed access links. 

We assume each processing stage of the VR application runs at AWS EC2~\cite{AWS_EC2} clusters tailored to its specific requirements:
\begin{itemize}
    \item Tracking: Uses C5 Amazon EC2 instances (c5.large) for its medium compute demands. 
    \item VR Processing: Utilizes C7g Amazon EC2 instances (c7g.xlarge) for its intensive compute tasks. 
    \item Rendering: Runs on G3 Amazon EC2 instances (g3s.xlarge), optimized for graphics. 
    \item Caching: Employs R7g Amazon EC2 instances (r7g.medium) for memory-intensive workloads. 
\end{itemize}

In addition, each access point hosts a general-purpose  cluster using M7i (m7i.large) instances, well-suited for diverse workloads that require a balanced combination of compute, memory, and networking resources. 

Table \ref{tab:network_config_scenC} details the resulting cloud-network resource capacities and cost values.

\begin{table}[h]
    \centering
    \begin{tabular}{|p{0.85in}|p{0.4in}|p{0.7in}|p{0.65in}|}
        \hline RESOURCE & MAX BLOCKS & CAPACITY per BLOCK & COST per BLOCK\\
        \hline
        \hline Core $\leftrightarrow$ Edge & $100$ & $50$ Mbps & $0.45$ \$/hr \\
        \hline Edge $\leftrightarrow$ Edge & $100$ & $50$ Mbps & $0.45$ \$/hr\\
        \hline Edge $\leftrightarrow$ Access & $100$ & $50$ Mbps & $0.90$ \$/hr\\
        \hline Tracking Cluster & $500$ & $3.4 \times 2$ $\times$ $16$ Gflops & $0.085$ \$/hr\\
        \hline VR Proc. Cluster & $500$ & $2.6\times 4 \times 64$ Gflops& $0.150$ \$/hr\\
        \hline Rendering Cluster & $500$ & $4.825$ Tflops & $0.750$ \$/hr \\       
        \hline Caching Cluster & $500$ & $8$ GiB & $0.054$ \$/hr\\        
        \hline General Purpose Cluster & $500$ & $2.6\times 2 \times 8$ Gflops  & $0.108$ \$/hr\\      
        \hline
    \end{tabular}
    \caption{\small{Cloud-network capacity and cost parameters (Scen. 2).}}
    \label{tab:network_config_scenC}
\end{table}

\subsubsection{\textbf{Service configuration}}
The VR service graph comprises three connected components, each representing different VR applications consumed by different user groups. Figs \ref{fig:og_scen2} and \ref{fig:tg_scen2} show the original service DAG and its corresponding DAG-to-Forest transformation, respectively.

We use Unreal Engine 5.3 specifications as a reference for VR application systems, with their recommended hardware requirements~\cite{unreal_requirement} listed in Table \ref{tab:unrealrequirement}. 
The associated VR service function resource requirements are then listed in Table \ref{tab:Func_req_scenarioC}. 
Additionally, Table \ref{tab:XRserviceconfig} describes the resulting production, communication, and consumption rates associated with each commodity. 

\begin{table}[h]
    \centering
    \begin{tabular}{|p{1in}|p{2in}|} 
        \hline \textbf{Processor} & Quad-core Intel or AMD, \textbf{2.5 GHz} \\ 

        \hline \textbf{Graphics Card} & DirectX 11 or 12 compatible graphics card with the latest drivers, such as GeForxe RTX 3060, \textbf{12.74 Tflops} \\ 
        
        \hline \textbf{Memory} & 8 GB RAM \\         
        \hline
    \end{tabular}
    \caption{\small{Unreal 5.3 Hardware requirements.}}
    \label{tab:unrealrequirement}
\end{table}


\begin{table}[h]
    \centering
    \begin{tabular}{|p{0.64in}|p{0.68in}|p{0.68in}|p{0.66in}|}
    \hline
    FUNCTION & CPU (Gflops) & GPU  (Gflops) & Memory (MB)\\
    \hline
    Tracking & $0.625$ & N.A. & $0.65$  \\
    \hline
    Content Store & N.A. & N.A. & $14000$ \\
    \hline
    Cache & $0.025$ & N.A. & $2800$ \\
    \hline
    VR Proc. & $1.75$ & $2548$ & $800$ \\
    \hline
    Rendering & $0.625$ & $8918$ & $148$ \\
    \hline
    \end{tabular}
\caption{\small{VR service function requirements for Scenario 2. 
}}
\label{tab:Func_req_scenarioC}
\end{table}

\begin{table}[h]
    \centering
        \begin{tabular}{|p{0.98in}|p{0.58in}|p{0.58in}|p{0.50in}|}
        \hline Commodity & Comm. Rate (Mbps) & Prod. Rate (Gflops) & Cons. Rate (MB) \\
        \hline (gNB, Tracking) & $1$ & N.A. & $0.65$ \\
         \hline (Tracking, VR\_Proc) & $17.28$ & $0.625$ & $800$ \\
        \hline (Data store, Cache) & $2$ & N.A. & $2800$ \\
        \hline (Cache, VR\_Proc) & $50$ & $0.025$ & $800$ \\
        \hline (VR\_Proc, Render) & $120$ & $2549.75$ & $148$ \\
        \hline (Render, gNB) & $80$ & $8918.63$ & N.A. \\
        \hline
        \end{tabular}
    \caption{\small{VR service commodity rates.}}
    \label{tab:XRserviceconfig}
\end{table}


In terms of latency, we set the propagation delay of Core$\leftrightarrow$Edge, Edge$\leftrightarrow$Edge, and Edge$\leftrightarrow$Access to 20 ms, 15 ms, and 10 ms, respectively. We then impose a maximum service latency of 50 ms for the two top service components in Fig.~\ref{fig:tg_scen2}, and 150 ms for the bottom one.


\subsubsection{\textbf{Results}}

In this scenario, we use larger scaling factors representative of, not only increased service quality (e.g., higher resolution), but also high burstiness factors. 

\begin{figure}[h]
    \centering
    \includegraphics[width=\linewidth]{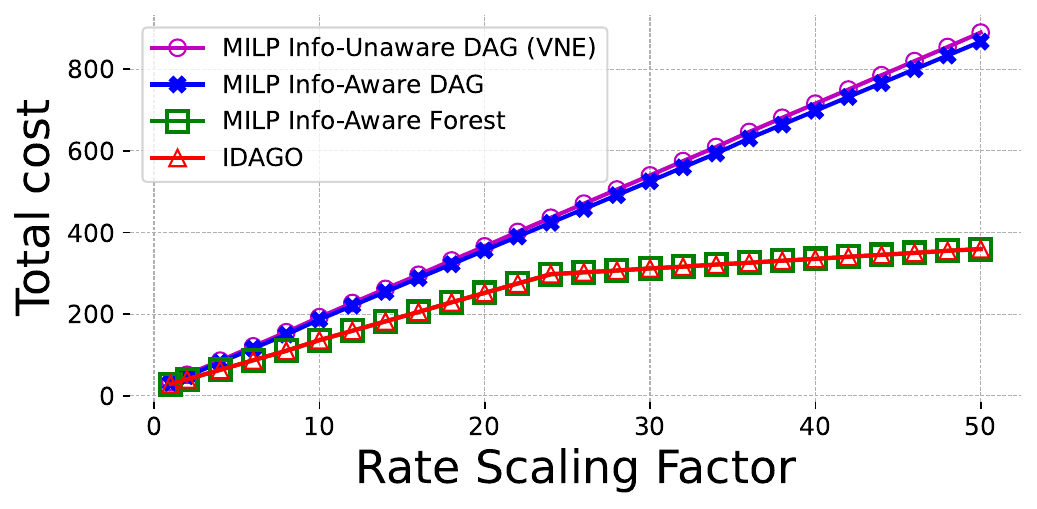}
    \caption{\small{Total resource cost vs rate scaling factor in Scenario 2.}}
    \label{fig:total_cost_scenC}
\end{figure}

The results in Fig.~\ref{fig:total_cost_scenC} confirm the insights observed in Scenario 1, in terms of the cost reductions obtained by the proposed information-aware service orchestration framework and associated \alg\, algorithm. 
In this scenario, characterized by realistic cloud-network and service parameters, \alg\, is able to achieve cost reductions of up to $2.5$X when compared to
state-of-the-art solutions that cannot effectively exploit information-awareness, as shown in
Fig.~\ref{fig:total_cost_scenC}. 
Additionally, \alg\, also reduces E2E service latency compared to both MILP Info-Unaware DAG and MILP Info-Aware DAG. Specifically, for the latency-tolerant service (bottom service in Fig.~\ref{fig:tg_scen2}), \alg\, achieves the same latency, and for the two latency-sensitive services, it achieves 35\% lower latency.



\section{Conclusion}
In this paper, we thoroughly investigate the optimal end-to-end orchestration of emerging NextG services over distributed cloud-networks. We show how the unique properties of NextG media services, including their 
complex processing graph structures and opportunities for data sharing and replication, disrupt the suitability of traditional VNE-based approaches. We propose a CNFlow based formulation that effectively captures the essential flow chaining, scaling, and replication characteristics of NextG media services. We then design \alg\, (Information-Aware DAG Orchestration), the first polynomial-time multi-criteria approximation algorithm for this challenging problem class that includes existing service orchestration problems as a special case. \alg\, leverages a novel DAG-to-Forest graph transformation procedure and existing information-unaware tree embedding techniques to provide polynomial-time solutions to the information-aware DAG orchestration problem with approximation guarantees. 
Extensive simulations in the context of NextG media services validate \alg's effectiveness, demonstrating substantial performance improvements over even exponential-time state-of-the-art approaches. 

\section*{Acknowledgements}
This work was partially supported by the European Union under the Italian National Recovery and Resilience Plan (NRRP) of NextGenerationEU, partnership on "Telecommunications of the Future" (PE00000001 - program "RESTART"), by the PRIN project "Resilient delivery of real-time interactive services over NextG compute-dense mobile networks" (E53D2300055000), and by funds from the US National Science Foundation as specified in the RINGS program.


\bibliographystyle{IEEEtran}
\bibliography{references_v2}

\setcounter{page}{1}
{\appendices
\section{Proof of Lemma \ref{decomp}}
\label{app_lemmadecomp}
First, we consider the validity of the embeddings obtained by the \alg \, algorithm. For each embedding, Algorithm \ref{disjoint} ends up mapping every function and commodity of each  service tree, $\mathcal R^{T,\phi} \in \mathcal R^T$, onto one node and one path in the cloud-augmented graph $\mathcal G^a$, satisfying Definition \ref{validmapping}.
More precisely, the single destination function $j \in \mathcal I^{d, T,\phi}$ is mapped to the only physical node $d^{\mathcal I}(j) \in \mathcal V^d$ (line 8), i.e., the node where the placement of $j\in \mathcal I^{d, T,\phi}$ was initialized to via constraint \eqref{eq:dest}. Similarly, since the placement of each source function  $i \in \mathcal I^{s,T,\phi}$ is set, via constraint \eqref{eq:source}, by the LP solution, into the only physical node $u=s^{\mathcal I}(i) \in \mathcal V^s$ that it is supposed to host it, Algorithm \ref{disjoint},  in Procedure 1  line 3,   maps each source function  $i \in \mathcal I^{s,T,\phi}$ to its fixed location $u=s^{\mathcal I}(i) \in \mathcal V^s$.
Algorithm \ref{disjoint} updates the set of current mapped functions $\mathcal Q$ every time a new function is mapped (line 13) and it runs until the set is empty (line 9), removing a function to each iteration (line 10). Hence, each function node is mapped exactly once. 
For a given commodity $k=(i, j) \in \mathcal K^{T,\phi}$, Procedure 1 can only map function $i$ after function $j$ has been mapped. Function $i$ can only be mapped to physical node $v \in \mathcal{V}^a$ if there is a path from a node $v$ to the already mapped function $j$ (that is $\map^{\phi, {\mathcal I}}_{n}(j)$). Accordingly, the commodity mappings are valid since every commodity $k=(i, j) \in \mathcal K^{T,\phi}$ is always mapped to a path $\mathcal P \in \mathcal G^a$ starting at either the already mapped location of $i$ (in case $i$ is a source function) or a valid location of $i$ and ending at the already mapped location of $j$. $\mathcal P$ is a path whose edges have positive flow of commodity $k$. Finally, since each service functions and commodity will be eventually mapped, the overall embedding is valid \cite{Poularakis2020Mobihoc}.

Next, we show that the decomposition of valid embeddings is complete, i.e. that $\sum_n p^\phi_n=1, \forall n\in \mathcal{N}$. 
In fact, in 
Algorithm \ref{disjoint}:
\begin{itemize}
\item[i)] At iteration $n$, $\fLP^{\phi}_{\rightarrow{d}}$
 denotes the (residual)
flow consumed, at that iteration, by the single destination function, $j \in 
\mathcal{I}^{d,T,\phi}$; 
\item [ii)] At $n=0$ $\fLP^{\phi}_{\rightarrow{d}}$ starts with a value of 1;
\item[iii)] At each iteration, $n >0$, its value is decreased by $p_n$;
\item[iv)] at any iteration $n >0$, $ 0 < p_n \leq \fLP^{\phi}_{\rightarrow{d}}$;
\item[v)] the  iterations  continue as long as $\fLP^{\phi}_{\rightarrow{d}} > 0$.
\end{itemize}
Therefore $1 \leq \sum_n p^\phi_n \leq 1$.
Since Algorithm \ref{disjoint} stops when $\fLP^{\phi}_{\rightarrow{d}} = 0$, in order to complete the proof of Lemma \ref{decomp}, it is enough to prove that,  in $\mathcal G^a$,  at the last iteration (after line 18)   no  residual commodity flows  of the restriction of the LP solution to  service component $\mathcal{R}^{T,\phi} $, are different from zero. This follows immediately from the facts  
 that in addition to the fractional commodity flow variables $\{\fLP^k_{uv}\}$, the fractional residual commodity flows computed in Line 18 of Algorithm~\ref{disjoint} and the binary embedding commodity flows, $\flD^k_{uv}(\Emb^{\phi}_n)$, computed in Step 4 of Algorithm~\ref{disjoint} during the $n$-th rounding try, also satisfy the generalized flow conservation constraints \eqref{eq:flowcons}.

\section{Proof of Lemma \ref{lemmaexpinf}}
\label{app_lemmaexpinf}
Since $R^{k}=R^{o}, \forall k \in g^{-1}_T(o)$, then
 the Random Information Flow for object $o$ over the link $(u,v) \in \mathcal{E}^{a}$, defined in Definition \ref{randomvariables2} can be reformulated as follow:
$$\displaystyle \muR_{uv}^o = {\max_{k \in g_T^{-1}(o)} \left \{\flR_{uv}^k R^k\right\}} =  R^o {\max_{k \in g_T^{-1}(o)}\left \{ \flR_{uv}^k \right\} }.$$

Therefore, using Lemma \ref{lemmaexpCom}, $\muR_{uv}^o(D)$  is a binary random variable taking values in the set $\{0,R^{o}\}$ with probability 
$$\mathbb{P}(\muR_{uv}^o = 0)= \prod_{k\in g_T^{-1}(o)} (1-\fLP_{uv}^k
),$$
and, obviously, $\mathbb{P}(\muR_{uv}^o = R^o)= 1-\mathbb{P}(\muR_{uv}^o = 0)$.

\section{Proof of Theorems \ref{thm1.1}, \ref{thm1.2}, \ref{thm1.3}}
\label{app_thm1}
Before starting the proof of Theorems \ref{thm1.1}, \ref{thm1.2}, \ref{thm1.3}, we recall the Hoeffding inequality statement.

\begin{proposition}
\label{pro:Hoeffding}
    (Hoeffding Inequality). Let $X = \sum_{i=i}^n X_i$, $X_i \in [a_i,b_i]$ be a sum of $n$ independent random variables. The following holds for any $t\geq 0$:
    $$
    \mathbb{P}(X - \mathbb{E}(X) \geq t) \leq \exp\bigg(- \frac{2t^2}{\sum_{i=1}^n (b_i - a_i)^2} \bigg)
    $$
\end{proposition}

Using Proposition \ref{pro:Hoeffding}, and 
observing that $ \mathbb{P}(0 \leq \muR_{uv} \leq \xi_{uv})=1$,  the probability that the total information  flow, $\muR_{uv}$,   over the  edge $(u,v)\in \mathcal{E}^{a}$ exceeds capacity $c_{uv}$ by a factor $\Delta_{\beta_1}$,  can be bound as follows:
\begin{eqnarray}
\label{theh1eq}
 \mathbb{P}(  \muR_{uv}\geq \Delta_{\beta_1} c_{uv}) \! \!\!\! \! \!\!\! \! \!\!\!
 \! \!\!\!   \! \!\!\! \nonumber \\
 & &  \leq \mathbb{P}(  \muR_{uv} -\mathbb{E}[\muR_{uv}]\geq \beta_1 c_{uv})  \\
 &&  \leq  \exp\Bigg[-\frac{2(\beta_1 c_{uv})^2 }{\xi_{uv} }\Bigg] 
 \label{theh1eq2} 
\end{eqnarray}
where   \eqref{theh1eq2} follows from  Hoeffding's lemma.

Analogously, letting $\CostLP$  the objective function value of the optimal fractional solution,   we have that:

\begin{eqnarray}
     \mathbb{P}(\CostR \geq \Delta_{\alpha} \Cmilp) \! \!\!\! \! 
     \! \!\!\!   \! \!\!\! \nonumber \\
     & &  \leq 
     \mathbb{P}(\CostR \geq \Delta_{\alpha} \CostLP) \! \!\!\! \! \!\!\! \! \!\!\!
     \! \!\!\!   \! \!\!\! \label{eq:CmilClp} \\
     & &  \leq \mathbb{P}(  \CostR -\mathbb{E}[\CostR]\geq \alpha \CostLP)  \nonumber \\
     &&  \leq  \exp\Bigg[-\frac{2(\alpha  \CostLP)^2}{\chi}\Bigg]
     \label{eq:CmilClp1} 
\end{eqnarray}
where \eqref{eq:CmilClp} follows from the observation that $\sf \Cmilp \geq \CostLP$ while \eqref{eq:CmilClp1} follows 
from Proposition \ref{pro:Hoeffding} and from  the fact that
$ \mathbb{P}(0 \leq \CostR \leq \chi)=1$.

Finally, using the same machinery we can bound the probability the aggregate latency of the destination commodity $k \in \mathcal K^{d}$ violates the maximum latency $L^k$ by a factor $\Delta_{\beta_2}$,  as follows: 

\begin{eqnarray}
    \mathbb{P}(\latencyR^k_T\geq \Delta_{\beta_2} L^k) \! \!\!\! \! \!\!\! \! \!\!\!
     \! \!\!\!   \! \!\!\! \nonumber \\
     & &  \leq \mathbb{P}(  \CostR -\mathbb{E}[\latencyR^k_T]\geq \gamma L^k)  \nonumber \\
     &&  \leq  \exp\Bigg[-\frac{2(\gamma  L^k)^2}{(\Lambda^k_{\sf max}-\Lambda^k_{\sf min})^2 }\Bigg]
\label{imbecille}
\end{eqnarray}
where \eqref{imbecille} follows, immediately from Proposition \ref{pro:Hoeffding} and from the fact 
$ \mathbb{P}(0 \leq \latencyR^k_T \leq \Lambda^k_{\sf max})=1$.

\section{Proof of Theorem \ref{thm2}}
\label{app_thm2}

Denote by 
 $\mathcal{A}$ the event that 
the objective value $\CostR$ exceeds $\sf \Cmilp$ by a factor $\Delta_\alpha$.  Let $\mathcal{L}^k$ denote the event that the random aggregate latency of  destination commodity $k$, $\latencyR^k_T$, exceed $L^k$ by a factor $\Delta_{\beta_2}$.  Finally
denote by $\mathcal{B}_{uv} $ the event that  $\muR_{uv}$ violates the capacity of link $(u,v) \in \mathcal{E}^{a}$.
Then:

\begin{align}
\label{eqthe3}
\displaystyle 
P_{\alpha, \beta_1, \beta_2} & \triangleq
\mathbb{P}\left( \displaystyle  
\cup_{(u,v) \in \mathcal{E}_{LP}^{a}} \mathcal{B}_{u,v} \cup\mathcal{A} \cup_{k\in\mathcal{K}^d} \mathcal{L}^k \right )\\
&\leq \sum_{(u,v) \in \mathcal{E}^{a}}
\mathbb{P}\left(\mathcal{B}_{u,v}  \right)+
\mathbb{P}\left(\mathcal{A} \right) +
\sum_{k \in \mathcal{K}^d} \mathbb{P}\left(\mathcal{L}^k \right)  \\
& \!\!\!\!\!\!\!\!\!\!\!\!\!\!\! \leq |\mathcal{E}^{\sf F}|
e^{\Big[-\frac{2(\beta_1 c_{uv})^2 }{\xi_{uv}}\Big]}\!+\!
|\mathcal{K}^d| e^{\Big[-\frac{2(\beta_2 L^k)^2 }{\left(\Lambda^k_{\sf max} - \Lambda^k_{\sf min} \right)^2 }\Big]}
\!+\! e^{\Big[-\frac{2(\alpha \CostLP)^2 }{\chi}\Big]}
\end{align}
Therefore, the probability that after $t$ trials the \alg \, algorithm does not return an
$(\Delta_\alpha;\Delta_{\beta_1}, \Delta_{\beta_2})$-approximation with relaxed constraints  with high provability,  for the \prob\, problem in \eqref{eq:ilp}, is given by: 
\begin{align}\label{bi-criteria}
\left(
|\mathcal{E}^{\sf F}|
e^{\Big[-\frac{2(\beta_1 c_{uv})^2 }{\xi_{uv}}\Big]}\!+\!
|\mathcal{K}^d| e^{\Big[-\frac{2(\beta_2 L^k)^2 }{\left(\Lambda^k_{\sf max} - \Lambda^k_{\sf min} \right)^2 }\Big]}
\!+\! e^{\Big[-\frac{2(\alpha \CostLP)^2 }{\chi}\Big]}
\right)^t
\end{align}
which goes to zero as $t$ increase if:
$$
|\mathcal{E}^{\sf F}|
e^{\Big[-\frac{2(\beta_1 c_{uv})^2 }{\xi_{uv}}\Big]}\!+\!
|\mathcal{K}^d| e^{\Big[-\frac{2(\beta_2 L^k)^2 }{\left(\Lambda^k_{\sf max} - \Lambda^k_{\sf min} \right)^2 }\Big]}
\!+\! e^{\Big[-\frac{2(\alpha \CostLP)^2 }{\chi}\Big]} = \theta,
$$ 
with $\theta < 1$.
To this end, we impose that:
\begin{eqnarray}
\frac{\theta}{3} &=& |\mathcal{E}^{\sf F}|e^{\Big[-\frac{2(\beta_1 c_{uv})^2 }{\xi_{uv}}\Big]} \label{eq:beta1}\\
&=& |\mathcal{K}^d| e^{\Big[-\frac{2(\beta_2 L^k)^2 }{\left(\Lambda^k_{\sf max} - \Lambda^k_{\sf min} \right)^2 }\Big]}  \\
&=& e^{\Big[-\frac{2(\alpha \CostLP)^2 }{\chi}\Big]}.
\end{eqnarray}
Starting from \eqref{eq:beta1}, using \eqref{eq:rottaenorme}, we have:
\begin{eqnarray}
 e^{\Big[-\frac{2(\beta_1 c_{uv})^2 }{\xi_{uv}}\Big]} \leq \exp\Bigg[-\frac{2(\beta_1 c_{min})^2 }{\xi_{\sf max}}\Bigg] 
\leq \frac{1}{\frac{\theta}{3}|\mathcal{E}^{\sf F}|},   
\end{eqnarray}
from which it follows:
\begin{eqnarray}\label{eq:beta1end}
\beta_1 \geq \sqrt{\frac{1}{2} \ln{\frac{\theta}{3} |\mathcal{E}^{\sf F}|}}\frac{\sqrt{\xi_{\sf max}}}{c_{\sf min}}.    
\end{eqnarray}
Using \eqref{eq:beta1end} and letting $R_{\sf max} = \max_{o \in \mathcal O} \{R^o\}$, from the feasibility condition $ R_{\sf max} \leq  c_{uv}$ for all $(u,v) \in \mathcal E^a$, we also have: 
$$
\beta_1 \geq \kappa \sqrt{\frac{1}{2} \ln{\frac{\theta}{3} |\mathcal{E}^{\sf F}|}}.
$$
with $0\leq\kappa \leq 1$.

Furthermore, by definition, we have that:  
$$
\max_{(u,v)\in \mathcal{E}^{\sf F}} \left\{ \delta_{\beta_1}  \right\} =
\max_{(u,v)\in \mathcal{E}^{\sf F}} \left\{  \frac{\mathbb{E}[\muR_{uv}]}{c_{uv}}
\right\} 
$$


Therefore,
we have that  $\Delta_{\beta_1}$ needs to satisfy:
\begin{align}\label{eq:proof_thm2}
    \Delta_{\beta_1} \geq \kappa \sqrt{\frac{1}{2}\ln \frac{\theta}{3} |\mathcal{E}^{\sf F}|} 
     + \max_{(u,v)\in \mathcal{E}^{\sf F}} \left\{  \frac{\mathbb{E}[\muR_{uv}]}{c_{uv}}
\right\}.
\end{align}

Analogously, following similar steps, we have that $\Delta_{\beta_2}$ and $\Delta_\alpha$ can be chosen as:
\begin{eqnarray}
\Delta_{\beta_2} = \sqrt{\frac{1}{2}\ln (\frac{\theta}{3} |\mathcal{K}^d)|}  \frac{(\Lambda_{\sf max}-\Lambda_{\sf min}) }{(L_{min })} + \delta_{\beta_2}
\end{eqnarray}
\begin{eqnarray}
    \Delta_\alpha = \frac{1}{\CostLP}\sqrt{\chi\frac{\ln(\frac{\theta}{3})}{2}} + \delta_\alpha,
\end{eqnarray}


\section{Proof of Corollary \ref{corrollary_resous_block}}
\label{app_corrollary_resous_block}

According to Theorem \ref{thm1.1}:

$$\mathbb{P}(  \muR_{uv}\geq \Delta_{\beta_1} c_{uv})  \leq  \exp\Bigg[-\frac{2(\beta_1 c_{uv})^2 }{\xi_{uv}}\Bigg].
$$
Using the fact that $c_{uv}$ can be written as $c_{uv}^T \cdot c_{uv}^b$, it follows that:
\begin{eqnarray}
\label{corollary_eq1}
 \mathbb{P}(  \muR_{uv}\geq \Delta_{\beta_1} c_{uv}) \! \!\!\! \! \!\!\! \! \!\!\!
 \! \!\!\!   \! \!\!\! \nonumber \\
 & &  = \mathbb{P}(\muR_{uv}\geq \Delta_{\beta_1} c_{uv}^T \cdot c_{uv}^b)  \\
 &&  = \mathbb{P}\big(  \frac{\muR_{uv}}{c_{uv}^b}\geq \Delta_{\beta_1} c^T_{uv}\big)\label{eq:corollary_eq2} \\
&& \leq  \exp\Bigg[-\frac{2(\beta_1 c_{uv})^2 }{\xi_{uv}}\Bigg]
\label{eq:corollary_eq3}
\end{eqnarray}
where \eqref{eq:corollary_eq3} follows from 
    Theorem \ref{thm1.1}. 
    Consequently, we want to determine that 
    \begin{align}\label{eq:corollary_thes}
    \mathbb{P}\bigg(\Bigl\lceil \frac{\muR_{uv}}{c_{uv}^b} \Bigr\rceil \geq \Delta_{\beta_1} c^T_{uv}\bigg),
    \end{align}
    knowing that $\Bigl\lceil \frac{\muR_{uv}}{c_{uv}^b} \Bigr\rceil$  is the smallest integer $n$ such that $ n\geq \frac{\muR_{uv}}{c_{uv}^b} $.

    Using the fact that  $\Bigl\lceil \frac{\muR_{uv}}{c_{uv}^b} \Bigr\rceil = \frac{\muR_{uv}}{c_{uv}^b} + b$ with $b$ 
    $0 \leq b < 1$, 
    we can rewrite Eq. \eqref{eq:corollary_thes} as follow:
    \begin{eqnarray}
     \mathbb{P}\bigg(\frac{\muR_{uv}}{c_{uv}^b} + b \geq \Delta_{\beta_1} c^T_{uv}\bigg) &\!\!\! \! \! = \!\!\! \! \!&  \mathbb{P}\bigg(\frac{\muR_{uv}}{c_{uv}^b} + b \geq \Delta_{\beta_1} c^T_{uv}\bigg)\!\!\! \! \! \\
     &  \!\!\! \! \!& \mathbb{P}\bigg(\frac{\muR_{uv}}{c_{uv}^b}  \geq \Delta_{\beta_1} c^T_{uv} - b\bigg)  \\
     &  \!\!\! \! \!\leq \!\!\! \! \!&  \mathbb{P}\bigg(\frac{\muR_{uv}}{c_{uv}^b} \geq \Delta_{\beta_1} c^T_{uv} - 1 \bigg) \label{eq:bound_corollary} \\
     & \!\!\! \! \!\leq \!\!\! \! \!& \exp\Bigg[-\frac{2(\beta_1 c_{uv}^T -1)^2 }{\xi_{uv}^T}\Bigg], \label{eq:proof_coroll}
    \end{eqnarray}
where $\xi_{uv}^T = \frac{\xi_{uv}}{c_{uv}^b}$.  and  \eqref{eq:proof_coroll} follows from Theorem \ref{thm1.1}.

\section{Notations}
\begin{table*}[]
    \centering
    \begin{tabular}{|p{1.7in}|p{5in}|}
    \hline
    Notation & Description \\
    \hline
    $\mathcal G = (\mathcal V, \mathcal E)$; $\mathcal G^a = (\mathcal V^a, \mathcal E^a)$ & Network graph, associated nodes ($\mathcal V$) and links ($\mathcal E$); Cloud-augmented graph, associated nodes ($\mathcal V^a$) and links ($\mathcal E^a$).\\

    
    $\mathcal V^p; \mathcal V^s; \mathcal V^d$ & Computation nodes; Source nodes; Destination nodes.\\

    $\mathcal E^c; \mathcal E^p; \mathcal E^s; \mathcal E^d$ & Communication links; Computation links; Source links; Destination links.\\

    $\mathcal E^{p-}; \mathcal E^{p+}$ & Computation in links (storage resources); Computation out links (processing resources).\\

    $\mathcal N^{-}(u); \mathcal N^{+}(u)$ & Incoming and outgoing neighbors of node $u\in\mathcal V^a$.\\

    $c_{uv}; w_{uv}$ & Capacity and cost of link $(u,v)$.\\
    
    $\mathcal R = (\mathcal I, \mathcal K)$ & Information-aware service (or service collection) graph, composed of functions ($\mathcal I$) and commodities ($\mathcal K$).\\
    
    $\mathcal I^s; \mathcal I^d; \mathcal I^p$ & Source functions; Destination functions; Computation functions.\\
    
    $\mathcal K^s; \mathcal K^d; \mathcal K^p$ & Source commodities; Destination commodities; Processing commodities. \\


        
     $\mathcal X(k)$ & Set of input commodities required to produce commodity $k \in \mathcal K$.\\
     
    $s^{\mathcal K}(k); d^{\mathcal K}(k)$ & Source node hosting the function producing commodity $k \in \mathcal K^s$; Destination node hosting the function consuming commodity $k \in \mathcal K^d$.\\
    
        
    $s^{\mathcal I}(i); d^{\mathcal I}(i)$ & Node hosting source function $i \in \mathcal I^s$; Node hosting destination function $i \in \mathcal I^d$.\\


    $\mathcal V^{p,\mathcal I}(i)\equiv\mathcal V^{p,\mathcal K}(k)$ & Computation nodes that can host function $i$ and hence produce commodity $k$.\\

    $R^k_{uv}$ & Rate of commodity $k \in \mathcal K$ when it goes over link $(u,v) \in \mathcal E^a$.\\

    $\mathcal O$; \quad $g:\mathcal K \rightarrow O$ & Set of information objects; Information mapping function.\\

    $f_{uv}^k; \mu_{uv}^o; \mu_{uv}$ & Virtual commodity flow,  actual information object flow, and actual information flow variables of MILP \eqref{eq:ilp}.\\    

    $l^k_{uv}; l^k; l^k_T; L^k$ & Latency to transmit or process a unit of commodity $k$ over link $(u,v)$; Local latency of commodity $k$; Cumulative latency of commodity $k$; Maximum service latency associated with destination commodity $k$.\\

    $c^T_{uv}; c^b_{uv}; w^b_{uv}; y_{uv}; \burfact$ & Total number of blocks; Capacity per block; cost per block; Allocated blocks; Burstiness factor.\\

    
    \hline
    
    $\mathcal R^T = (\mathcal I^T, \mathcal K^T)$ & Transformed service forest and associated functions and commodities. \\ 
    $\mathcal{K}^{s, T}$ & Source commodities of $\mathcal R^T$. \\
    $\mathcal{X}^{T}(k)$ & Set of input commodities required to produce commodity $k \in \mathcal K^T$. \\
    $\mathcal R^{T,\phi} = (\mathcal I^{T,\phi}, \mathcal K^{T,\phi})$ & $\phi-th$ connected component (tree) of transformed service graph (forest) $\mathcal R^T$.\\

    $\mathcal I^{s,T,\phi}; \mathcal I^{d,T,\phi}$ & Source functions and destination functions of $\mathcal R^{T,\phi}$.\\

    $\mathcal K^{s,T,\phi}; \mathcal K^{d,T,\phi}$ & Source commodities and destination commodities of $\mathcal R^{T,\phi}$.\\


    $g_T:\mathcal K^T \rightarrow O$ & Information mapping function for transformed service graph $\mathcal R^T$.\\

    $M=|\mathcal K^d|$ & Number of connected components (service trees) in transformed service graph $\mathcal R^T$.\\

    $\fLP^k_{uv}; \muLP^o_{uv}; \muLP_{uv}$ & Fractional commodity flow, object flow, and information flow solution from LP relaxation of MILP \eqref{eq:ilp}. \\

    $\fLP^{\phi}_{\rightarrow{d}}$ & Fractional commodity flow consumed by the single destination function of $\mathcal R^{T,\phi}$ throughout the Decomposition step of \alg.\\

    $\Emb^{\phi}_{n} = (m_n^{\mathcal I,\phi}, m_n^{\mathcal K, \phi})$ & Embedding, composed of function mapping and commodity mapping of service tree $\phi$ computed at iteration $n$ of the Decomposition step of \alg.\\

    $\miD^{\phi}= \left  \{ (\Emb^{\phi}_{1}, p^\phi_1) \ldots  (\Emb^{\phi}_{N_\phi}, p^\phi_{N_\phi}) \right \}$ & Decomposition of a service tree $\phi$, composed of a set of valid embeddings and associated probabilities.\\

    $N_\phi$ & Number of embeddings of service tree $\phi$ computed by \alg.\\

    $\flD^k_{uv}; \muD^o_{uv}; \muD_{uv}$ & Commodity flow, object flow, and information flow variables computed by \alg.\\
    
    
    $\underline{\Emb}=[\Emb^{1}, \Emb^{2}, \ldots, \Emb^{\nCp}]$ & Embedding of service collection $\mathcal R^T$ computed by \alg.\\

    $\ratio; \quad \nwuse$ & Cost Approximation Ratio; Capacity Relaxation Factor.\\
    
    \hline

    $\Cmilp; \CostLP; \CostIDAGO$ & Optimal objective function value of MILP~\eqref{eq:ilp}, LP relaxation and of \alg.\\

    $\Delta_\alpha; \Delta_{\beta_1}; \Delta_{\beta_2}$ & Approximation factor for optimal objective function value; Capacity Relaxation factor; Latency Relaxation factor.\\
    
    
    $\delta_\alpha; \delta_{\beta_1}; \delta_{\beta_2}$ & Ratio between the  expected value of $\CostR$ and $\CostLP$; Ratio between the expected value of $\muR_{uv}$ and $c_{uv}$; Ratio between the expected value of $\latencyR^k_T$ and $L^k$.\\
    

    $\alpha; \beta_1; \beta_2$ & $\Delta_\alpha - \delta_\alpha$;\quad $\Delta_{\beta_1} - \delta_{\beta_1}$;\quad $\Delta_{\beta_2} - \delta_{\beta_2}$. \\
    
    $\flR^k_{uv}; \muR^o_{uv}; \muR_{uv}$ & Random Commodity Flow; Random Information Object Flow; Random Information Flow.\\

    $\latencyR^k; \latencyR^k_T; \CostR$ & Random Local Latency; Random Cumulative Latency. Random Resource Cost.\\
    

    $\yLP; \yD; \gamma_{uv}$ & Resource allocation solution from LP relaxation of MILP \eqref{eq:ilp}; Resource allocation computed by \alg; Random Resource Blocks.\\


    

    









    \hline
    \end{tabular}
    \caption{Main notations.}
    \label{tab:notations_extended}
\end{table*}

}
\end{document}